\documentclass{ws-ijbc}
\usepackage{ws-rotating} 
\begin{document}

\catchline{}{}{}{}{} 

\markboth{Katsanikas and Patsis}{The structure of invariant tori in a 3D
    galactic potential}

\title{The structure of invariant tori in a 3D
    galactic potential}
\author{M. KATSANIKAS}
\address{Research Center for Astronomy, Academy of Athens\\
  Soranou Efessiou 4,  GR-11527 Athens, Greece}
\address{Section of Astrophysics, Astronomy and Mechanics, \\Department of
  Physics, University of Athens, Greece\\ mkatsan@academyofathens.gr}
\author{P.A. PATSIS}
\address{Research Center for Astronomy, Academy of Athens\\
  Soranou Efessiou 4,  GR-11527 Athens, Greece\\patsis@academyofathens.gr }

\maketitle

\begin{history}
\received{(to be inserted by publisher)}
\end{history}

\begin{abstract}
We study in detail the structure of phase space in the neighborhood of stable
periodic orbits in a rotating 3D potential of galactic type. We have used the
color and rotation method to investigate the properties of the invariant tori
in the 4D spaces of section. We compare our results with those of previous
works and we describe the morphology of the rotational, as well as of the tube
tori in the 4D space. We find sticky chaotic orbits in the immediate
neighborhood of sets of invariant tori surrounding 3D stable periodic
orbits. Particularly useful for galactic dynamics is the behavior of chaotic
orbits trapped for long time between 4D invariant tori. We find that they
support during this time the same structure as the quasi-periodic orbits
around the stable periodic orbits, contributing however to a local increase of
the dispersion of velocities. Finally we find that the tube tori do not appear
in the 3D projections of the spaces of section in the axisymmetric Hamiltonian
we examined. 
\end{abstract}

\keywords{Chaos and Dynamical Systems, Galactic Dynamics, 4D surfaces of 
section, Invariant tori}

\twocolumn{
\section{Introduction}
The method of surfaces of section for the study of dynamical systems
dates back to Poincar\'{e} (1892) and has many applications to
Dynamical Astronomy (for a review see e.g. Contopoulos 2002). A basic
problem in Hamiltonian Systems of three degrees of freedom is the
visualization of the 4D\footnote{throughout the paper we will refer to
  ``n-dimensional'' as ``nD''; i.e. 3D, 4D etc.} surfaces of section.
Let us assume the phase space of an autonomous Hamiltonian system,
that has 6 dimensions, e.g. in Cartesian coordinates, $(x,y,z,\dot
x,\dot y,\dot z)$.  For a given value of the Hamiltonian a trajectory lies on
a 5D 
manifold.  In this manifold a surface of section is 4D.  This does
not allow us to visualize it directly. Several
methods have been applied to overcome this problem in the past and we
summarize them below.

The structure of the 4D surfaces of section was examined for the
first time in the pioneer work of Froeschl\'{e} (1970, 1972). In
that work he used stereoscopic views and the method of slices in order
to understand the structure of tori, that appeared at the neighborhood of
stable periodic orbits. Similar methods have been applied by Martinet
and Magnenat (1981), Contopoulos et al. (1982), and Magnenat (1982) for
studying the 3D projections of invariant tori in the 4D surface of
section or in the phase space of a 4D symplectic map. The 2D
projections of such invariant tori have been examined on various 2D planes
in detail (e.g. Skokos, Contopoulos and Polymilis 1997, 1999).

In the present paper we use the method of color and rotation proposed
by Patsis and Zachilas (1994). In this method we first consider 3D
projections and rotate the 3D figures on a computer screen to observe
the figure from all its sides. Then we use colors to indicate the 4th
dimension. For this purpose we make use of the ``Mathematica'' package
(Wolfram 1999) and of its subroutines. Each point is colored according
to the value of its 4th coordinate in the following way: Firstly we
define the surface of section that we will use, e.g. $y=0$ with $\dot
y >0$. Secondly we select a 3D subspace of the surface of section,
e.g.  $( x,\dot x,\dot z)$ and we determine the minimum and maximum
values of the 4th coordinate $z$. Finally we normalize the resulting
interval [min($z$), max($z$)] into [0,1] from which interval the color
functions of ``Mathematica'' take values. In our figures we give
always the color function of ``Mathematica'' that we used in a
color-coded bar. The point of view of the observer of the figures is given in spherical
coordinates. This specifies the direction along which the figure is
projected. The unit for distance $d$ of the consequents of the surface of 
section from the observer  is given by ``Mathematica'' in a special scaled 
coordinate  system in which the longest side of the bounding box, which 
surrounds the figure, has length 1. For all figures we use $d=1$. The method 
associates the smooth  distribution or the mixing of colors, with specific 
types of dynamical behavior in the 4th dimension (Patsis and Zachilas 1994).

In order to study the structure of phase space at the neighborhood of
a periodic orbit (p.o.), we first locate it by means of an iterative
method and calculate its stability.  The calculation of the linear
stability of a periodic orbit is based on the method of Broucke (1969)
and Hadjidemetriou (1975).  We first consider small deviations from
its initial conditions and then integrate the orbit again to the next
upward intersection. In this way a 4D map (Poincar\'{e} map) is
established, which is unique (Abraham and Marsden 1978 p. 521) and
relates the initial with the final point. The relation of the final
deviations of this neighboring orbit from the periodic one, with the
initially introduced deviations can be written in vector form as $\xi$
= M $\xi_{0}$. Here $\xi$ is the final deviation, $\xi_{0}$ is the
initial deviation $M$ is a $4\times 4$ matrix, called the monodromy
matrix. This matrix satisfies the symplectic identity and the map is
called symplectic (Arnold and Givental 2000). It can be shown that the
characteristic equation can be written in the form $\lambda^4 + a
\lambda^3 + \beta \lambda^2 + a \lambda +1 = 0 $. Its solutions
$\lambda_{i},\; i=1,2,3,4$, due to the symplectic identity of the
monodromy matrix, that obey the relations $\lambda_{1} \lambda_{2}=1$
and $\lambda_{3} \lambda_{4}=1$ can be written as

\begin{eqnarray} 
\lambda_1, \frac{1}{\lambda_1} = \frac {- b_1 \pm \sqrt{b_1^2 - 4}}{2}
\nonumber\\
\lambda_3, \frac{1}{\lambda_3} = \frac {- b_2 \pm \sqrt{b_2^2 - 4}}{2}
\nonumber\\
\end{eqnarray}
where

\begin{equation}
b_{1, 2} = \frac {a \pm \sqrt{\Delta}} {2}
\end{equation}

and

\begin{equation}
\Delta = a^2 - 4 (\beta - 2)
\end{equation}

The quantities $b_{1}$ and $b_{2}$ are called the stability indices.
Following the notation of Contopoulos and Magnenat (1985) if $\Delta > 0,\; |b_1| < 2$ and $|b_2| < 2$, all four
eigenvalues are complex on the unit circle and the periodic orbit is
called "stable" (S). If $\Delta > 0$ and $|b_1| > 2, \; |b_2| < 2$ or
$|b_1| < 2,\;|b_2| > 2$, the periodic orbit is called "simple
unstable" (U). In this case two eigenvalues are on the real axis and
two are complex on the unit circle. If $\Delta > 0$ and $|b_1| > 2$ and $|b_2|
> 2$, the periodic orbit is called "double unstable" (DU) and the four
eigenvalues are on the real axis. Finally if $\Delta < 0$ the periodic
orbit is called ``complex unstable'' ($\Delta$).
In this case the four eigenvalues are complex numbers and they are off
the unit circle. For the generalization of this kind of instability in
Hamiltonian systems of N degrees of freedom the reader may refer to
Skokos (2001). When two eigenvalues collide at $(1,0)$ of the unit
circle the parent family becomes simple unstable (U) and a new family of p.o. is born.  In this paper we examine the
evolution of the phase space at the transition from stability to simple instability ($S \rightarrow U$). The parent family in our example has an orbital plane,
i.e. it is 2D, and by becoming simple unstable as the energy
increases, it generates by bifurcation a stable 3D family of p.o..

According to the KAM theorem (Kolmogorov 1954, Moser 1962, Arnold
1963) in an almost Integrable Hamiltonian system of N degrees of freedom there are orbits that lie on
N-dimensional tori. The KAM theorem has been extended to
2n-dimensional almost integrable symplectic maps by Wiggins (2003,
p.225) and by Kuksin and P\"{o}schel (1994). This means that in a
2n-dimensional almost integrable symplectic map there are orbits that
lie on n-dimensional tori. In our 3D Hamiltonian system a 4D
Poincar\'{e} map is defined on the surface of section. In specific
cases this 4D Poincar\'{e} map is almost integrable. Then, according to
Wiggins (2003, p.225), we have orbits that lie on 2D invariant tori in
the 4D space of section. The structure of these tori, is the subject of our present, rather descriptive, paper.

In order to facilitate the discussion of the figures in the paper, we
give some useful definitions related to a torus. In the 3D
space, a torus is a surface that is generated when we rotate a circle on the x-z (or y-z) plane
around the axis $z$ (Fig. \ref{tor}). The definition of the angles $u$ and
$\upsilon$ are also noted in Fig.~\ref{tor}. The internal surface of the 
torus is defined as the set of points of the torus, where we have
$90^{o} \leq u \leq 270^{o}$.  The external surface of torus is
defined as the set of points of the torus, where we have $0^{o} \leq u
\leq 90^{o}$ and $270^{o} \leq u \leq 360^{o}$. We will call ``tori''
the objects we study in this paper, despite the fact that they seem to
be generated by an ellipse instead of a circle i.e. they are elliptic
tori.

\begin{figure}[h]
\begin{center}
\begin{tabular}{cc}
\resizebox{80mm}{!}{\includegraphics{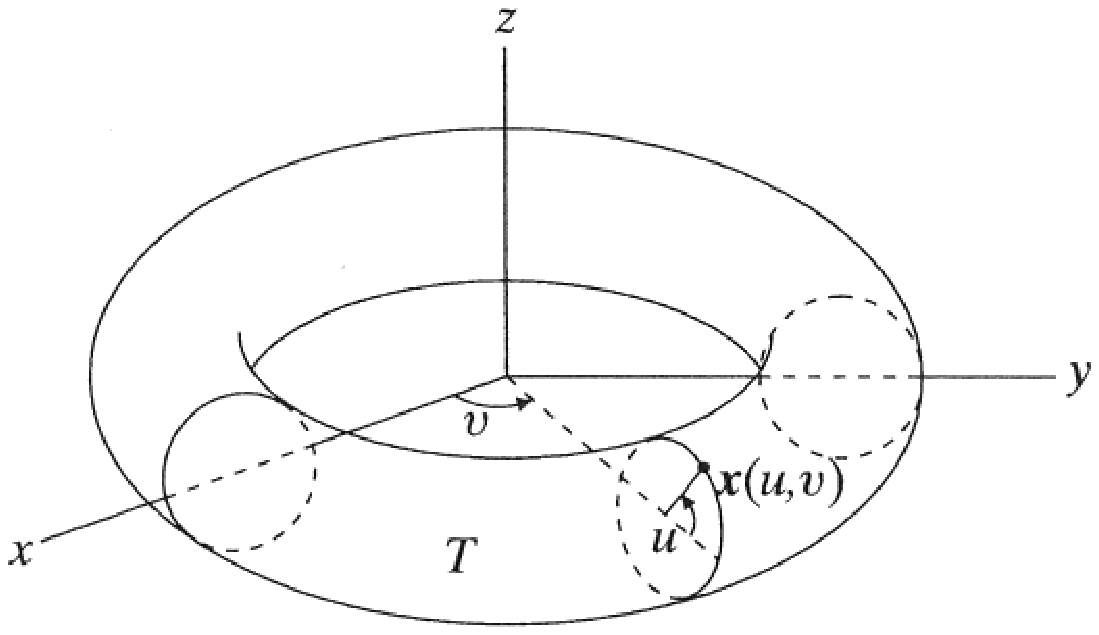}}\\
\end{tabular}
\caption{The definition of the $\upsilon$ and u angles on a torus.
Arrows indicate the direction along which the angles u,$\upsilon$ increase
from zero towards larger values.}
\label{tor}
\end{center}
\end{figure}

The purpose of this paper is to study the structure of the phase space in the neighborhood of stable periodic orbits in 3D galactic potentials. We want to understand how the phase space is structured and how it evolves as the stability of the x1, 2D central family of our system on the equatorial plane, changes from stable to simple unstable. We want also to understand how the phase space structure evolves as the main parameters of our system vary. In section 2 we describe our
Hamiltonian system, while in section 3 we present the morphological evolution of the families of periodic orbits involved in our study. Section 4 discusses the structure of the phase space when we perturb the initial conditions of our stable periodic orbits in the x-direction. We study successively the dynamical behavior close to the periodic orbit before (Sect.~4.1) and after (Sect.~4.2 ) the transition of the central family from stability to instability. We discuss as well the ``rotation numbers'' we define on the tori we have found (Sect.~4.3). Then, we increase the energy and we describe the changes we observe in the spaces of section in  Sect.~4,4. The perturbations in the z-direction are presented in Sect.~5, which has a similar structure as Sect.~4. In Sect.~6 we vary the perturbation of our system. Finally in section 7 we discuss our results and we enumerate
our conclusions.

\section{The  Hamiltonian System}
The  system   we  use  for  our  applications  rotates  around its  
z-axis  with  angular  velocity  $\Omega_b$ . The  Hamiltonian  of  the  
system in Cartesian coordinates  is :

\begin{eqnarray}
H(x, y, z, \dot x, \dot y, \dot z)=
\nonumber\\ 
\frac {1}{2}(\dot x^2 + \dot y^2 + \dot z^2) + \Phi(x, y, z)
\nonumber\\
- \frac{1}{2} \Omega_b^2 (x^2 + y^2)
\end{eqnarray}
 
where  $\Phi(x,y,z)$ is the potential we used in our
applications, i.e.:

\begin{eqnarray}
\Phi(x,y,z)= 
\nonumber\\
-\frac{GM_{1}}{(x^2+ \frac{y^2}{q_a^2} +[a_{1}+(\frac{z^2}{q_b^2}+
b_{1}^2)^{1/2}]^2)^{1/2}}-
\nonumber\\
 \frac{GM_{2}}{(x^2+ \frac{y^2}{q_a^2} +[a_{2}+(\frac{z^2}{q_b^2}+b_{2}^2)
^{1/2}]^2)^{1/2}}
\nonumber\\
\end{eqnarray}

The potential in it's axisymmetric form $(q_a = 1,q_b = 1)$ can de
considered as a representation of the potential for the Milky Way
approximated by two Miyamoto disks with masses $M_1$ and $M_2$
respectively (Miyamoto and Nagai 1975).  In our units, distance $R$=1
corresponds to 1 kpc. The velocity unit corresponds to 209.64~$km/sec$.
 For the Jacobi constant (hereafter called the
``energy'') Ej=1 corresponds to 43950 $(km/sec)^2$.We have used the
following values for the parameters: $ a_{1}=0 \; kpc,\; b_{1}=0.495
\; kpc,\; M_{1}=2.05 \times 10^{10} \; M_{\odot},\; a_{2}=7.258 \;
kpc,\; b_{2}=0.520 \; kpc,\; M_{2}=25.47 \times 10^{10} \;
M_{\odot},\; q_a = 1.2,\; q_b = 0.9$ and $\Omega_b = 60 \; km \;
s^{-1} \; kpc^{-1}$. The parameters $q_a, q_b$ determine the geometry
of the disks, while $a, b$ are scaling factors (Binney \& Tremaine
2008, p.73-74). The chosen $\Omega_b$ value puts corotation at
R=$4kpc$ and has been used by Englmaier and Gerhard (1999).

In Fig.~\ref{zvc} we give the $(E_j,x)$ Zero
Velocity Curve ``ZVC'' (see e.g. Contopoulos 2002 p.391). It refers to
orbits on the equatorial plane z=0 and separates the regions where
motion is allowed from those where it is forbidden.

\begin{figure}[h]
\begin{center}
\begin{tabular}{cc}
\resizebox{90mm}{!}{\includegraphics{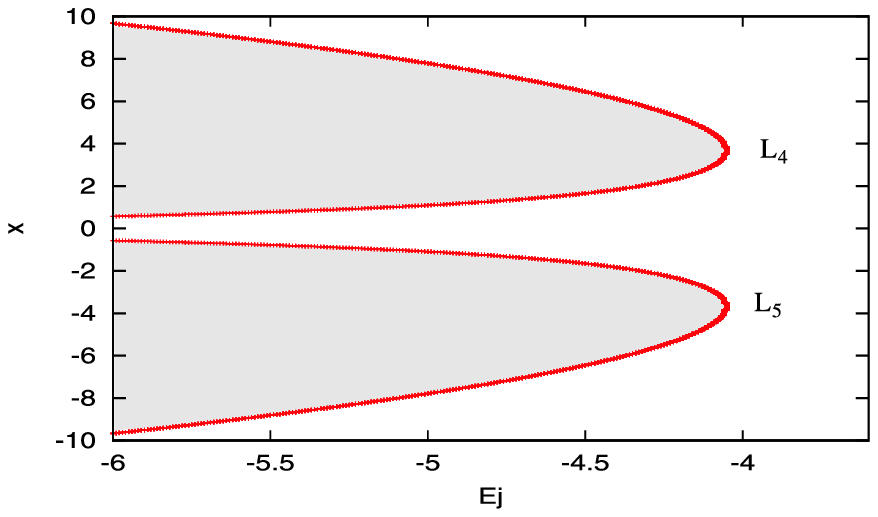}}\\
\end{tabular}
\caption{The (Ej,x) Zero Velocity Curve (ZVC), for orbits on the z =0
  plane, in our Hamiltonian system (4), with $\Phi$ as in (5) and  parameter 
  values given in the text. Motion is forbidden in the grey areas. L$_4$ and L$_5$ are the Lagrangian points on the minor axis of the galaxy.}
\label{zvc}
\end{center}
\end{figure}

\section{The orbital evolution along a $S \rightarrow U$ transition}

\begin{figure}[t]
\begin{center}
\begin{tabular}{cc}
\resizebox{85mm}{!}{\includegraphics{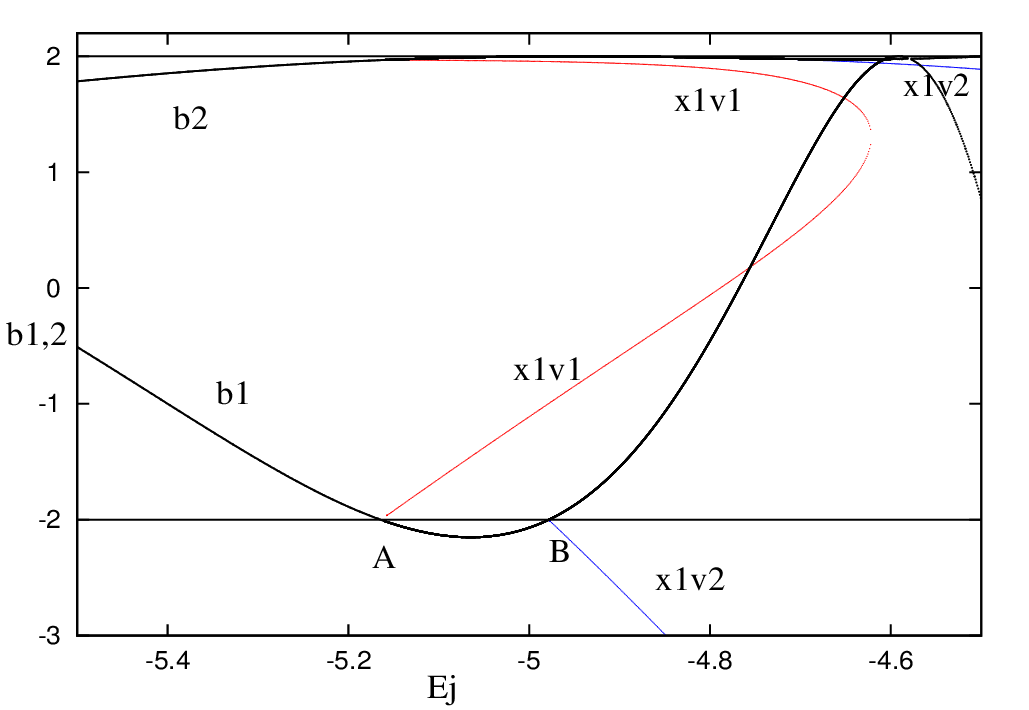}}\\
\end{tabular}
\caption{Stability diagram  for $-5.5 <$ Ej $< -4.5$, that shows the stability of the
  family x1 (black lines for its b1,b2 indices) and its bifurcating families of p.o. 
  x1v1 (red) and x1v2 (blue). The two indices of the family x1v1 join at Ej=$-4.62$ and the
  family becomes complex unstable.}
\label{stab1}
\end{center}
\end{figure}

A method to follow the stability of a family of periodic orbits in a
system is by means of the "stability diagram" (Contopoulos \& Barbanis 1985; Pfenniger 1985a). 
The stability diagram gives the evolution of the stability of a family of 
periodic orbits in a system as one parameter varies, by means of the evolution of the stability
indices b1, b2. In our case the parameter that varies is the energy
Ej. Fig.~\ref{stab1} gives the evolution of the stability of the
central family of periodic orbits in our system, x1 (Contopoulos and
Papayannopoulos 1980), and its bifurcations at the interval $-5.5 <$Ej$ <
-4.5$. We observe that x1 (black lines for its indices b1 and b2) is initially stable and at
Ej=$-5.1644$ it becomes simple unstable. There we have a $S
\rightarrow U$ transition and a new family, x1v1 (red lines), is
bifurcated and is stable. We call the transition point ``A''. The family x1
becomes stable again at Ej = $-$4.98, point ``B'', where we have an $U
\rightarrow S$ transition for x1. At ``B'', is introduced the
family x1v2 (blue line), which is initially simple unstable.

\begin{figure}[t]
\begin{center}
\begin{tabular}{cc}
\hspace{-7mm}
\resizebox{35mm} {!}{\includegraphics{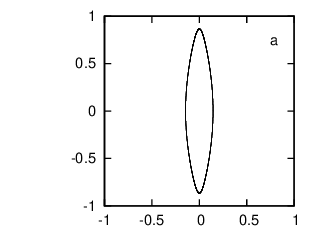}}
\hspace{-10mm}
\resizebox{35mm} {!}{\includegraphics{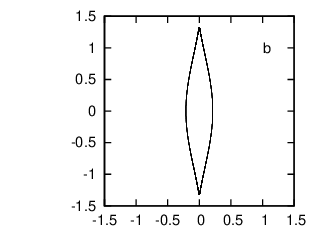}}
\hspace{-10mm}
\resizebox{35mm} {!}{\includegraphics{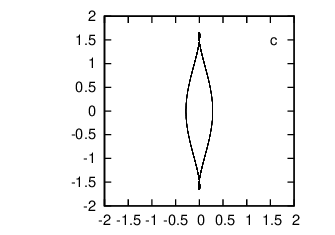}}\\
\end{tabular}
\caption{Periodic orbits of x1 at a) Ej= $-$5.624726 (before ``A''), 
b) Ej= $-$5.010526 (between ``A'' and ``B'') and c) Ej= $-$4.732626 (beyond
``B'').}
\label{orbx1}
\end{center}
\end{figure}

\begin{figure}
\begin{center}
\begin{tabular}{cc}
\hspace{-7mm}
\resizebox{35mm} {!}{\includegraphics{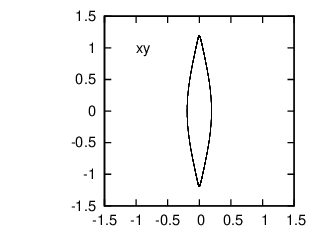}}
\hspace{-10mm}
\resizebox{35mm} {!}{\includegraphics{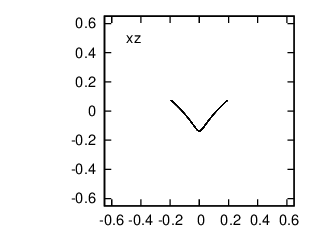}}
\hspace{-10mm}
\resizebox{35mm} {!}{\includegraphics{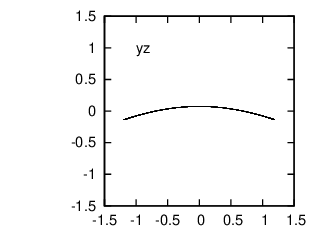}}\\
\end{tabular}
\caption{A typical orbit of x1v1  at Ej = $-$5.125377 (between
  ``A'' and ``B''). The $(x,z)$  and $(y,z)$
  projections  are  given  in  enlarged  scale, in order to better view the 
  corresponding morphology.}
\label{orbv1}
\end{center}
\end{figure}

In Fig.~\ref{orbx1} we present the morphological evolution of x1 as it
appears before ``A'', between ``A'' and ``B'', and beyond ``B''.  Point
``A'' is associated with the vertical 2/1 resonance (for a definition
see e.g. Contopoulos 2002 p. 379). The importance of the 3D
bifurcating family introduced at the vertical 2/1 resonance for the structure of galactic disks has been
underlined in several studies (Pfenniger 1985b; Patsis and Grosb{\o}l 1996; Skokos et al
2002a,b; Patsis et al 2002). Here we adopt the nomenclature of Skokos
et al 2002a,b and we call the bifurcating families
at Ej = $-$5.1644 and Ej = $-$4.98 x1v1 and x1v2 respectively. In
Fig.~\ref{orbv1} we give the morphology of x1v1 in a typical orbit at
energies between ``A'' and ``B''.

\section{Perturbations parallel to the equatorial plane}
\subsection{Spaces of section before the $S \rightarrow U$ transition}
First we examine the surfaces of section for our system at energies 
before ``A''. Figs.~\ref{2d}a and ~\ref{2d}b describe 
the surface of section  for  Ej=$-$5.207 (before ``A''). Despite the fact 
that the system we  investigate is of galactic type, the main goal of the 
present study is to  understand  the structure of the phase space in various 
cases in a 3D  autonomous  Hamiltonian system. In that sense we integrate our 
orbits for  times  necessary  to obtain  a clear view  of the dynamical  
phenomenon  we  study, regardless  of the  physical meaning of the integration 
time interval,  which  can be more  than a Hubble time. In Fig.~\ref{2d}a we 
observe  the invariant curves around periodic orbits of two 2D families 
located on the  equatorial plane z=0, in the $(x, \dot x)$ space
with initial conditions $(x_0 + \Delta x_0,\dot x_0,z_0,\dot z_0)$ $=(0.18312784
+ \Delta x_0,0,0,0)$ with $\Delta x_0 =0.1,\dots,0.8$   and 
$(x_0 + \Delta x_0,\dot x_0,z_0,\dot z_0)$ $=(-0.59595941 + \Delta x,0,0,0)$ with 
$\Delta x_0 =0.1,\dots,0.4$. Every  invariant curve  consists of $10^3$ 
consequents. The diagram describes a 
typical situation in rotating galactic potentials. The stable periodic orbit 
for $x>0$ belongs to x1 (initial conditions $(x_0,\dot x_0,z_0,\dot z_0)$ 
$=(0.18312784,0,0,0)$) and the  stable periodic for  $x<0$ to the retrograde 
family x4 (see Contopoulos 2002, p.391), which has initial 
conditions  $(x_0,\dot x_0,z_0,\dot z_0)$ $=(-0.59595941,0,0,0)$. The extent of the 
invariant curves is limited by the ZVC. Varying  the initial condition 
$x_0$ above the upper and below the lower limit of the ZVC at this Ej, always 
considering the plane of section $y=0$ (cf. Fig.~\ref{zvc}), we find the 
expected  bell-type curves ($10^3$ points in  the surface of section), which are related with 
escape orbits, as found by Contopoulos \& Patsis (2006) (Fig.~\ref{2d}b).

\begin{figure}[h]
  \begin{center}
    \resizebox{90mm}{!}{\includegraphics{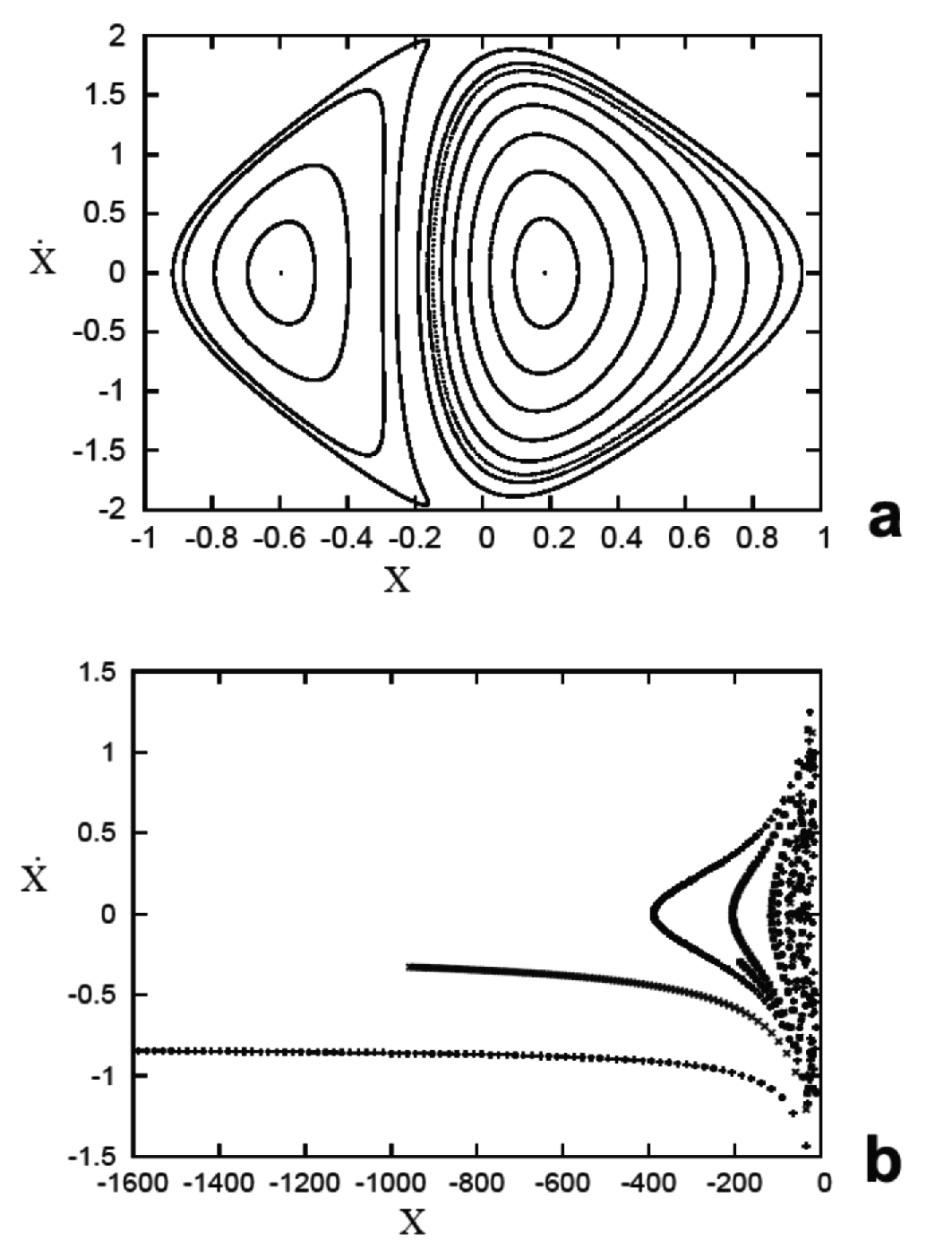}}
    \caption{The $(x, \dot x)$ surface of section for Ej=$-$5.207. (a) The invariant curves 
      are around the p.o. x1 (positive x) and x4 (negative x). (b) The surface of section considering initial conditions with $x$ above the upper and below the lower limit of the ZVC at this Ej. We observe the appearance of ``bell-type'' curves (Contopoulos \& Patsis 2006).}
    \label{2d}
  \end{center}
\end{figure}

\subsection{Spaces of section after the $S \rightarrow U$ transition}

Proceeding beyond ``A'', between ``A'' and ``B'', e.g. at Ej=$-$5.1574,
we encounter, always for the surface of section $y=0$, $\dot y >0$, two
simple periodic orbits with positive $x_0$. They are x1 (U), with initial conditions
$(x_0,\dot x_0,z,\dot z_0)$ =$(0.18958522,0,0,0)$ and x1v1 (S), with initial
conditions $(x_0,\dot x_0,z_0,\dot z_0) $ = $(0.18939859,0,0.030236585,0)$. We
investigate the phase space structure close to these two periodic
orbits, firstly by perturbing only the $x$ initial conditions of x1
and x1v1 by $\Delta x = 0.1,0.2\dots0.7$ successively. The 2D simple
unstable periodic orbit x1 lies on the equatorial $(x,y)$ plane of our
galactic model. By perturbing the initial conditions only in $x$ and
keeping the rest equal to 0, we encounter non-periodic orbits that
remain on the $(x,y)$ plane. Nevertheless, the surface of section of
our 3D autonomous Hamiltonian system is 4D, $(x_0, \dot x_0, z_0, \dot z_0)$,
and we can consider the $(x, \dot x,z)$ projection. For the simple unstable (U) periodic orbit x1, we have $z=0$
always, thus the $(x,\dot x, z)$ is identical with the $(x, \dot x)$
projection. This projection can be observed in Fig.~\ref{inc1}, where we have
seven invariant curves, surrounding the fixed point with the initial
conditions of x1. Each invariant curve has about $10^3$ consequents.
We name these curves $S_{1a}, S_{2a}, \dots ,S_{7a}$ for the $(x_0 +
\Delta x_0)$ perturbation of the $x_0$ initial condition with $\Delta x_0
=0.1,0.2,\dots0.7$ respectively.

\begin{figure*}[t]
  \begin{center}
   \resizebox{100mm}{!}{\includegraphics{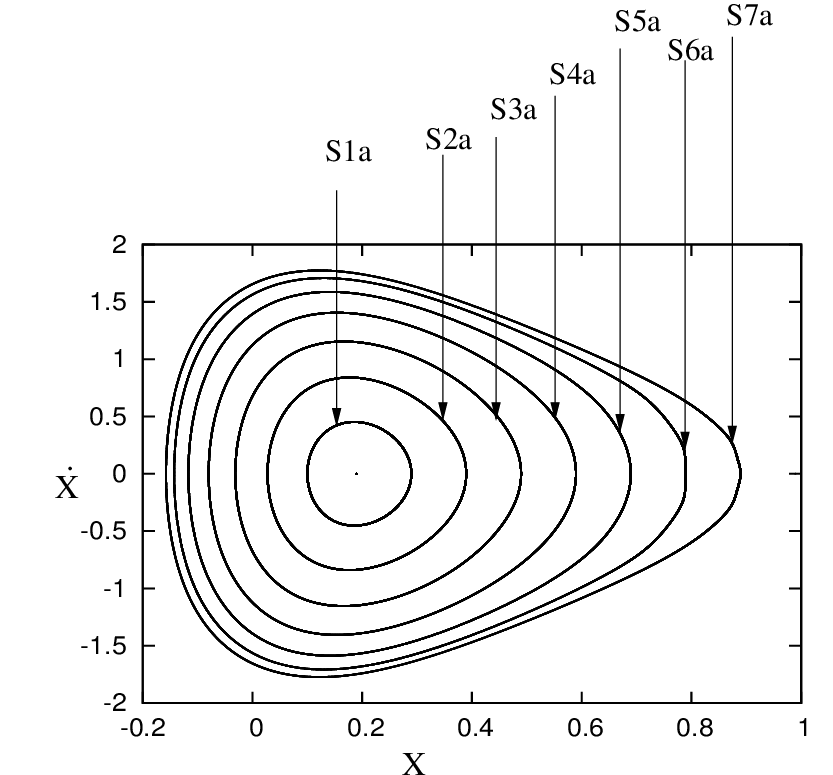}}
    \caption{The $(x, \dot x)$  projection of the 
      $(x, \dot x, z , \dot z)$ cross-section space at the  neighborhood  of 
      the 2D family x1 for $Ej=-5.1574$. The x1 periodic  
      orbit at $(0.18958522,0,0,0)$  is simple unstable. We
      name the invariant curves as $ S_{1a}, S_{2a}, S_{3a}, S_{4a}, S_{5a}, 
      S_{6a}, S_{7a}$ starting  with the one  closest to the periodic orbit.}
    \label{inc1}
  \end{center}
\end{figure*}

\begin{figure*}[t]
  \begin{center}
	\resizebox{110mm}{!}{\includegraphics{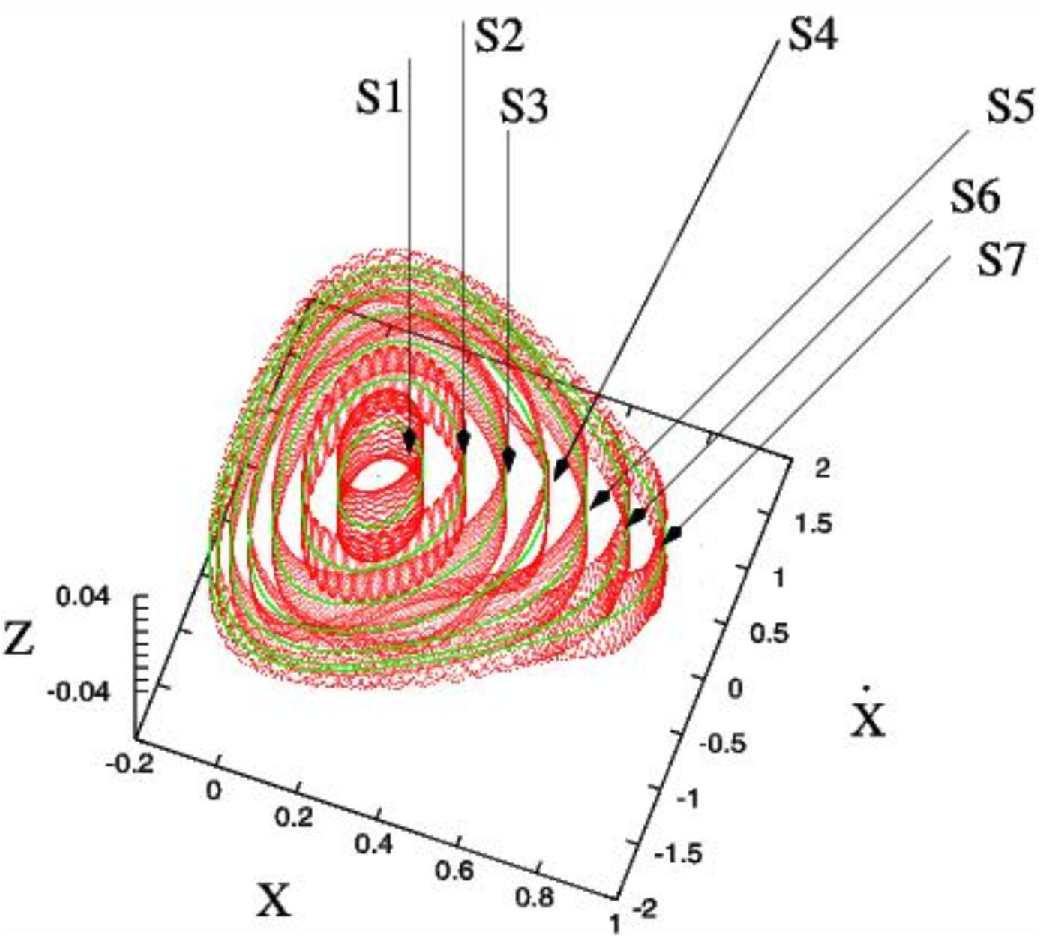}}
    \caption{ The $(x,\dot x,z)$ 3D projection of the $(x,\dot x,z,\dot z)$ 
      4D surface of  section at Ej=$-5.1574$. Our point of view is  $(\theta, \phi) = (17^{o}, 20^{o})$. 
      The $S_i$ tori around x1v1 (see text) are drawn  with red color, while the $S_{ia}$ 
      invariant curves around the x1 periodic are given with green 
      lines. }
    \label{int1}
  \end{center}
\end{figure*}

In a way this is a counterintuitive result, since  at this  Ej x1 is
characterized  as simple \textit{unstable}\footnote{Broucke (1969) 
characterizes this  type  of instability  as ``semi-instability'', 
subdividing  it  to even- and  odd- semi-instability,  depending  on whether  
the eigenvalues  on the real  axis are positive  or negative  respectively. In 
our case the eigenvalues are  positive (even- semi-instability).}, having two 
of its eigenvalues on the real  axis. We have also calculated the Lyapunov 
Characteristic Number (LCN) and we found it positive, equal to 0.025958. The
Lyapunov Characteristic Number (LCN)\footnote{When we use the term 
Lyapunov Characteristic Number we mean the maximal Lyapunov Characteristic 
Number.} of the  periodic orbits is defined as the maximum $\sigma_i$, where: 
$\sigma_i= \frac {1}{\tau} \ln(|\lambda_i|)$, $\tau$ is the period of the 
periodic orbit  and $\lambda_i$ the eigenvalues of the monodromy matrix of the 
Poincar\'{e}  map (e.g. Lichtenberg and Lieberman 1992  p.302, Skokos 2010). 
Figure~\ref{inc1}, underlines the fact, that  the  dynamical  behavior  close  
to a simple unstable  orbit  in  a 3D Hamiltonian  system  can be similar  to 
that  at the neighborhood  of a stable periodic orbit if the perturbation is 
restricted  in  one direction (in our case it is the radial one).

Applying the same seven  perturbations  to the $x_0$ initial 
conditions  of x1v1  we encounter in the  $(x, \dot x, z)$ space seven tori. 
The projection  of the  figure is given in Fig.~\ref{int1}. For this projection we use the point of view which is determined in spherical coordinates by the angles\footnote{The ($0^{o}, 0^{o}$) projection  brings the 
x-axis and y-axis  horizontally and vertically on the plane of the paper respectively and 
the z-axis perpendicular to them.} $(\theta, \phi) = (17^{o}, 20^{o}$) (see also introduction).
We observe seven tori drawn with red color surrounding  the x1v1  periodic orbit and we
name  them  $S_1, S_2, S_3, S_4, S_5, S_6$ and $S_7$ starting  with the 
closest  to  the  periodic one. We also plot with green  color the  $S_{1a}, 
S_{2a}, S_{3a}, S_{4a}, S_{5a}, S_{6a}$ and $S_{7a}$  invariant curves. Each 
torus  consists of $10^4$ consequents. Being just after the  bifurcation 
point, the  $x$  initial conditions of x1  and x1v1  are  almost identical. 
Thus, both  the  $S_i$ tori and the $S_{ia}$  invariant  curves  
surround  the ``common'' initial $x$ value. On the $(x,\dot x)$ projection the $S_i$ tori and the $S_{ia}$ invariant curves practically overlap.

\begin{figure}[h]
  \begin{center}
    \resizebox{90mm}{!}{\includegraphics{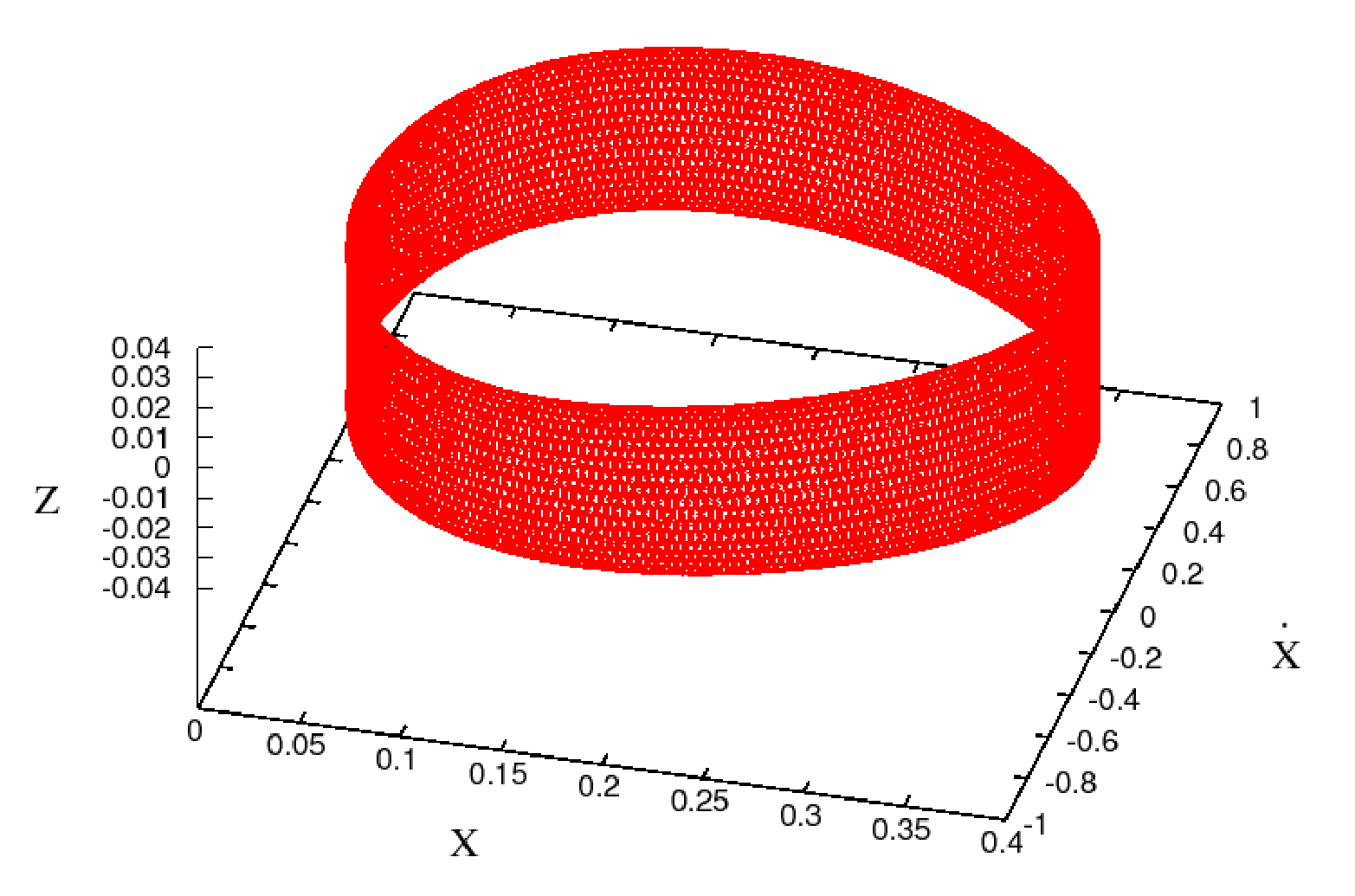}}
    \caption{The $(x, \dot x, z)$ projection of torus  $S_2$  in the 
       $(x,\dot x,z,\dot z)$  4D surface of section for Ej=$-5.1574$.   
       Our point of view in spherical  coordinates is now  
       $(\theta, \phi) = (40^{o},15^{o})$. This is a typical morphology of a 3D projection of a rotational torus.}
    \label{int2}
  \end{center}
\end{figure}

\begin{figure}
\begin{center}
\begin{tabular}{cc}
\resizebox{80mm} {!}{\includegraphics{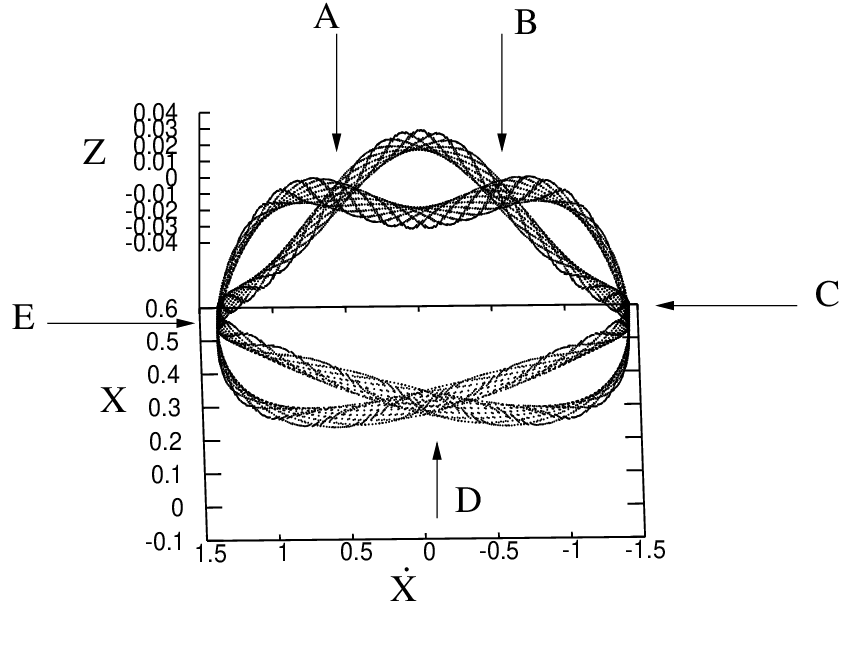}}\\
\end{tabular}
\caption{3D projection of torus $S_4$ from  the  point of  view 
  $(\theta, \phi) = ( 40^{o},264^{o})$. A,B,C,D and E are the five
  intersections of $S_4$  by itself. }
\label{projt}
\end{center}
\end{figure}

The $S_i$ tori are 3D projections of the invariant tori of our 4D
Poincar\'{e} map. Their structure, except that of $S_4$,
resembles the morphology of the objects defined as ``rotational tori''
by Vrahatis et al. (1996, 1997). These authors found the same kind of tori in a 4D
symplectic map related with the problem of beam stability in circular
particle accelerators. Even $S_2$, which seems to have its own
internal structure (Fig. ~\ref{int1}), becomes a typical rotational torus if we integrate
for time giving $4 \times 10^{5}$ consequents (Fig. ~\ref{int2}). An
exception from this morphology is the torus $S_4$, which has a thin,
complicated, ribbon-like structure. Fig.~\ref{projt} presents $S_4$
in the 3D $(x,\dot x,z)$ space. The figure helps us understand
that $S_4$ intersects itself at five places, namely A,B,C,D and E. In
practice this can be realized only by rotating the figure on the screen of our computer to understand its detailed morphology. The structure of our $S_4$ tori resembles the morphology of the objects defined as ``tube tori'' again
by Vrahatis et al. (1997).

Let us have  now a closer look at the 4D structure of the 
rotational $S_i$ tori by  applying  the method  of color and rotation. 
The tori occupy a subspace  of the 4D space of section bounded by  
$[x_1,x_2] \times [\dot x_1, \dot x_2] \times [z_1, z_2]
\times [\dot z_1, \dot z_2] = [-0.154927, 0.890256] \times [-1.79387,1.77590]
\times [-0.0302, 0.0302] \times $ $[-0.096, 0.096]$. 
As an example of the 4D structure of a rotational torus we depict in 
Fig.~\ref{rot1} the first  $4\times10^5$ consequents of the $S_2$ torus. 
This torus has a small thickness, as it is generated by rotation around the $\dot x$-axis of a thin ellipse.
We have chosen to plot the 
consequents  in  the $(\dot x,z, \dot z)$ projection and color them according 
to their values in the $x$-dimension. With the help of a graphic software we 
have rotated $S_2$ in our computer screen  in order to view it from all 
different perspectives and  better understand its internal structure. 
We observed that there is no mixing of colors on the surface of $S_2$. On the 
contrary, we find that the color variation on it follows some rules. 
This property of the ``rotational tori'' was already known by Patsis \& Zachilas (1994). 
However, the details, presented here for the first time, indicate a generic
behavior that characterizes this class of objects.

Moving along the  $\upsilon$-direction of the torus for constant $u$
(see Fig.~\ref{tor}) the  consequents keep their color by changing
from the external to the internal side of the torus and vice versa when they reach
certain four lines along the directions labeled with A, B, C and D in
Fig.~\ref{rot1}.

\begin{figure*}[t]
  \begin{center}
    \resizebox{70mm}{!}{\includegraphics{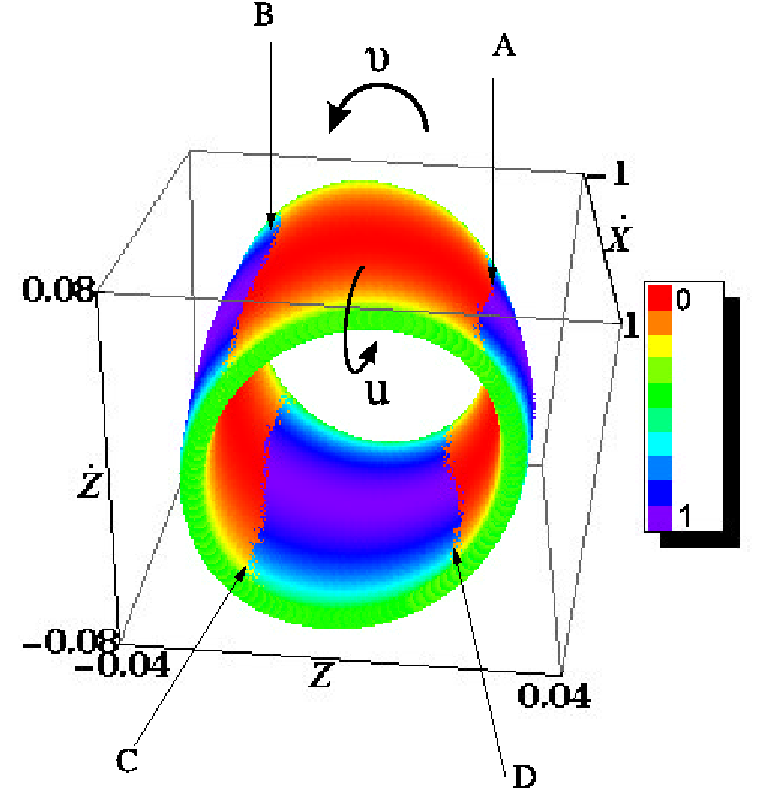}}
    \caption{ The torus $S_2$ in the  $(x,\dot x,z,\dot z)$ 4D surface of section 
      for Ej=$-5.1574$. The location of the consequents is given in the $(\dot x,z,\dot z)$ projection and are colored according to their value in the $x$ coordinate. Our point of view in  spherical  coordinates is $(\theta, \phi) = (180^{o},9^{o})$.}
    \label{rot1}
  \end{center}
\end{figure*}

\begin{figure*}
  \begin{center}
    \resizebox{110mm}{!}{\includegraphics{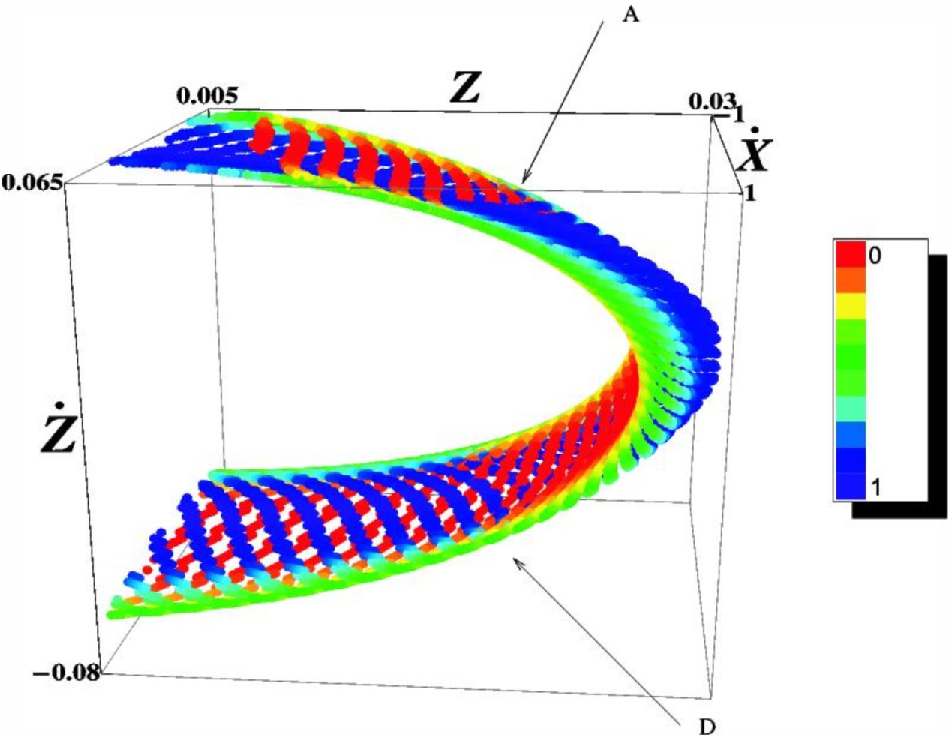}}
    \caption{The right part of the torus in Fig.~\ref{rot1} in enlargement. We observe how consequents of a certain color change from one side of the torus to the other.}
    \label{rot2}
  \end{center}
\end{figure*}

 The details of the color variation 
close to  these lines  as we move along 
the  $\upsilon$- direction for constant $u$, is shown in Fig.~\ref{rot2}. This
is an  enlargement of the  right part  of Fig.~\ref{rot1}, where
e.g. red  colored consequents on  the external side of the torus  change side sliding to 
the internal one, along the line indicated with  ``A'' (upper left side in  Fig.~\ref{rot2}). If we continue moving 
clockwise on the  $\upsilon$-direction the red consequents, now on the 
internal side of the  torus, continue until we reach  area  ``D''. There, they  
change  again  side  and  turn to the external side. This allows us very clearly  to 
observe the way  that the  transition  from one side to the other  happens. 
At the lower side of the  figure  we  can  observe how the red points slide 
behind the blue, forming a net. The ``red net'' is filled with red if we continue the calculation for longer time. Similarly the ``blue net'' on the inner side of the torus will be filled by blue after a long time and so do the 
red consequents at the lower left part of Fig.~\ref{rot2} will be covered
by the blue. In our example we started moving along the $\upsilon$-direction
from a red point. In general, the  succession of colors in the 
$\upsilon$-direction with constant $u$ angles  can be  understood by looking at 
the color-coded  bar  given at the  right of the figures with the colored tori, e.g. in Fig.\ref{rot1}. 
At the change of side of colors, red is combined with violet-blue (above or below it), orange  with blue, yellow with light-blue etc. and  
finally at $u=90^{o}$ and $u=270^{o}$ the consequents  have the shade of  
green (middle of the color-code bar).

Besides the color variation in the $\upsilon$-direction, there is
also color variation along $u$ (constant $\upsilon$).  If we examine
e.g. a ``red-dominated'' area on the $S_2$ torus in Fig.~\ref{rot1} we observe, that we get a
red ($x=0$) in the middle of the torus ($u=0$). At first for $u:0^{o}
\rightarrow 90^{o}$ we observe a color variation from red to yellow and then to green,
for values $\dot x=0 \rightarrow 1$ (Fig.~\ref{rot1}). Secondly for
$u:90^{o} \rightarrow 180^{o}$, at the internal side of the torus, the
colors change from green to light blue, then to blue and finally to violet (now for values $\dot x=1
\rightarrow 0$). Then, for $u:180^{o} \rightarrow 270^{o}$ the colors
change from violet to green (for values $\dot x=0 \rightarrow -1$) and
finally for $u:270^{o} \rightarrow 360^{o}$, at the external side of
the torus, we can observe a smooth color succession from green to yellow and then to red
for values $\dot x=-1 \rightarrow 0$. In conclusion we observe a
cyclic variation of colors (values of $x$) along the $u$-direction. This means that the color variation is also smooth along $\dot x$, as described above. The color variation is similar for all rotational $S_i$ tori.


\begin{figure*}[t]
  \begin{center}
\resizebox{120mm}{!}{\includegraphics{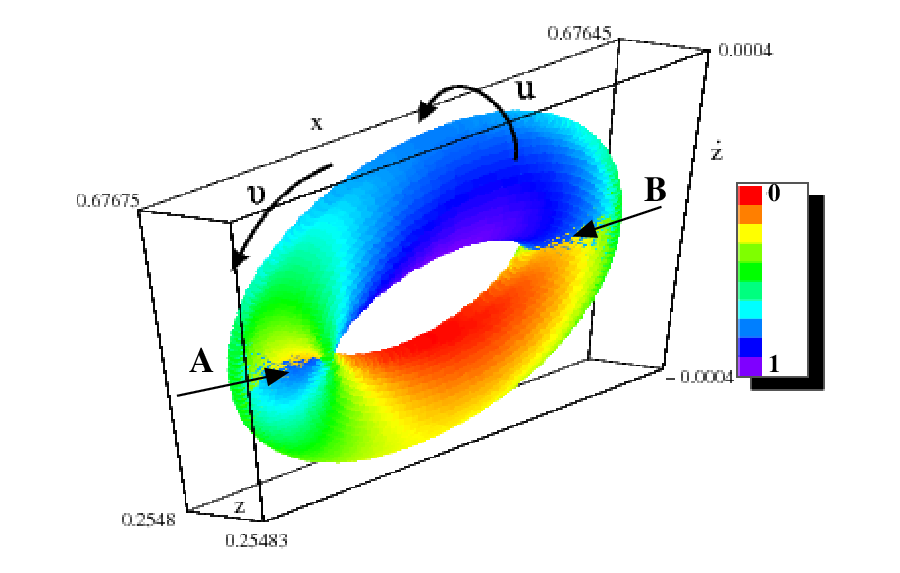}}
    \caption{The torus $S_a$ around a stable p.o. x1v3 in the  
      $(x,\dot x,z,\dot z)$  surface of section  for Ej=$-4.3$. The 
      consequents are colored according to their value  in the $\dot x$ 
      coordinate.  Our point of view in  spherical  coordinates is 
      $(\theta, \phi) = (30^{o},30^{o})$. We observe again a smooth color 
      variation on the surface of the torus. The color changes side along the lines indicated with arrows.}
    \label{donat1}
  \end{center}
\end{figure*}

\begin{figure*}
  \begin{center}
    \resizebox{110mm}{!}{\includegraphics{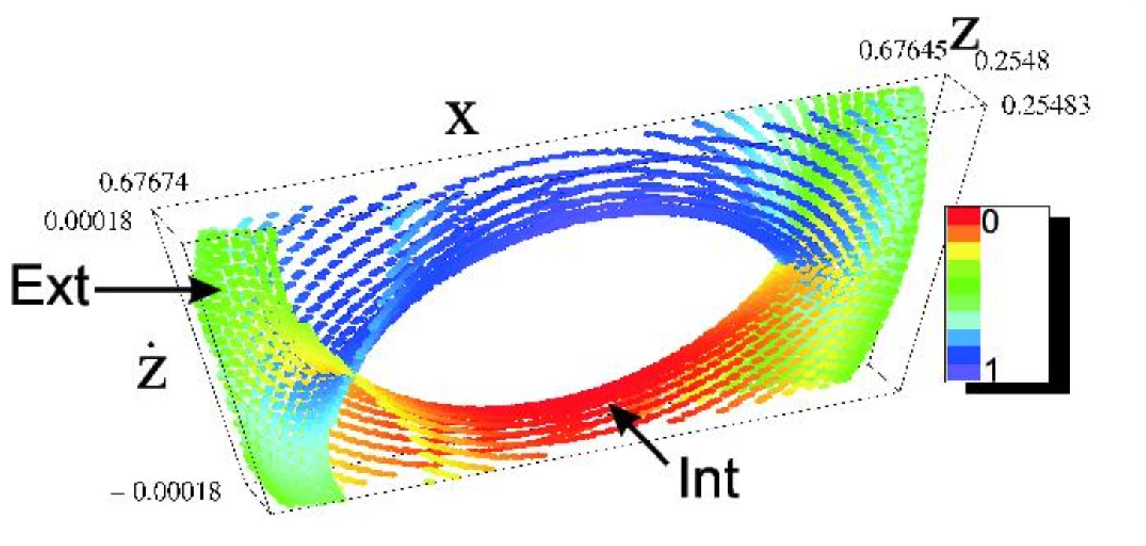}}
    \caption{The region of the intersection of the external surface with the
      internal surface of the torus $S_a$  in  the  $(x,\dot x,z,\dot z)$ 
      4D surface of section  for Ej=$-4.3$. The consequents are colored 
      according to their value  in the $\dot x$ coordinate.  
      Our point of view in  spherical  coordinates is   
      $(\theta, \phi) = (30^{o},30^{o})$. Arrows indicate the external and internal sides of the torus.}
    \label{donat1a}
  \end{center}
\end{figure*}

\begin{figure*}
 \begin{center}
   \resizebox{110mm}{!}{\includegraphics{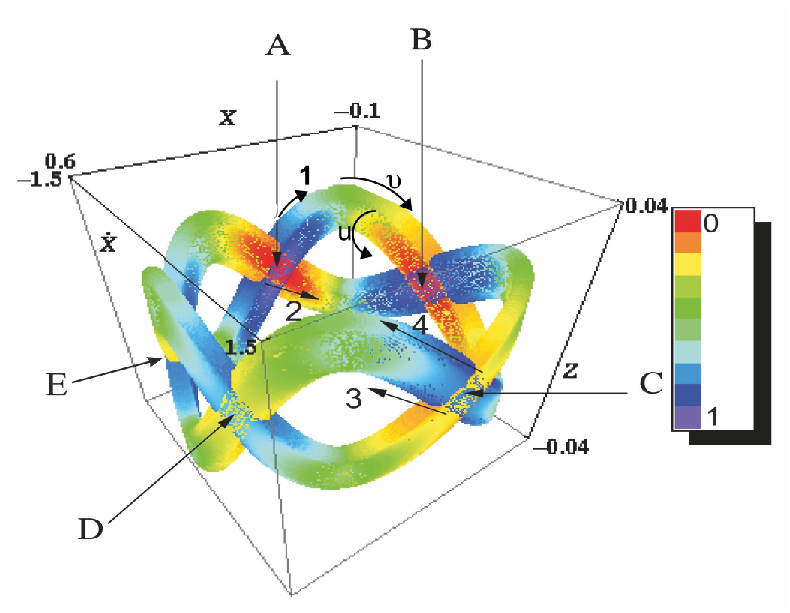}}
    \caption{ The $(x,\dot x,z,\dot z)$ 4D surface of section of torus $S_4$ 
      for Ej=$-5.1574$. The consequents are colored according to their value 
      in  the $\dot z$ coordinate.  Our view angles are 
      $(\theta, \phi) = (30^{o}, 45^{o})$. }
    \label{tub}
  \end{center}
\end{figure*}

 \begin{figure*}
   \begin{center}\resizebox{68mm}{!}{\includegraphics{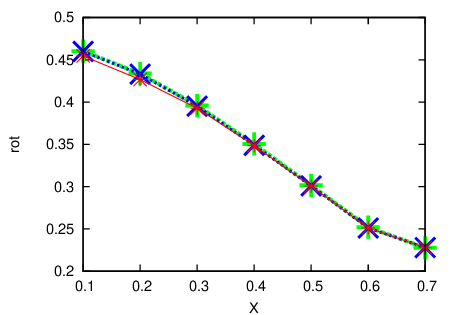}}
    \caption{Rotation curves at $Ej= -5.1574$ for the invariant curves around
      the periodic 
      orbit of the  x1 in the 2D subspace $(x,\dot x)$  (green line), around
      the periodic  
      orbit of the x1v1 in the 2D subspace $(x,\dot x)$ (blue line), and
      around the periodic  
      orbit of the x1v1  in the 3D subspace $(x, \dot x, z)$ (red line). We
      observe that the three curves have a very similar variation, and
      practically overlap.}
    \label{rcu1}
  \end{center}
\end{figure*}

In other cases of our system  we find elliptic tori with considerable smaller ellipticity.
We give an example not belonging to the $S\rightarrow U$ transition we use in our presentation.
By perturbing the $x$ initial condition
of a stable periodic orbit with initial conditions  $(x_0,\dot x_0,z,\dot z_0)$
=$(0.6765982,0,0.254816,0)$ for Ej=$-4.3$ (family x1v3,
Skokos et al. 2002a,b) with $\Delta x=10^{-5}$ we  observe a torus which we 
name $S_a$ (Fig.~\ref{donat1}). In Figs.~\ref{donat1} and \ref{donat1a} 
torus $S_a$ is depicted in the 3D subspace $(x,z,\dot z)$ 
of the 4D surface of section and is  colored according to  the values of the 
4th dimension $\dot x$. In Fig.~\ref{donat1} we observe that at the upper 
region of the external surface of the $S_a$ dominate blue shades. Along the 
$\upsilon$-direction  we  observe a smooth color variation from blue to 
light blue, to green and to yellow. At a region indicated with an arrow labeled ``A'', we observe
that yellow meets the blue. At this region we have an intersection of the 
external surface with the internal surface of the torus 
$S_a$. The details of the 4D structure of the torus close to the intersection is described in Fig. \ref{donat1a}. The color continues from a yellow shade on the external surface (indicated with ``Ext'') of the torus $S_a$ to orange and then to red on the internal  
surface of the torus (indicated with ``Int'' in Fig. \ref{donat1a}). We observe a smooth variation of 
colors from red to the orange on the internal surface of the torus $S_a$
(see also Fig. \ref{donat1}) until a region where orange meets the blue. At this region, indicated with arrow ``B'', 
we have in the 3D $(x,z,\dot z)$ projection again an intersection of the external surface of the $S_a$  with 
the internal surface of the torus $S_a$ along a line.
 
We want to apply now the color and rotation method to the other kind of tori we found around x1v1 in Fig.~\ref{int1}.
As we already remarked, $S_4$ has a different morphology than the rest of
the $S_i$'s, i.e. it is a tube torus. In the 3D 
projection $(x,\dot x,z)$ (Fig.~\ref{projt}) we have  realized that  $S_4$, intersects itself in  
five regions (A, B, C, D and E). In Fig. \ref{tub} the $S_4$ torus is colored according to  the 
$\dot z$ values. In this figure  we observe a  smooth color variation at the 
areas between two successive  intersections.  For example 
we examine the area between the intersections A and B on the internal surface 
of the torus along the arrow ``1''.  If we move  from  A as indicated with the 
arrow``1'' (Fig. \ref{tub}), for constant $u$, and starting with
blue colored points, we can see  a smooth  color variation  from blue to light 
blue, to green, to yellow, to orange 
at the intersection
B. Now if we  move from the intersection  A  along the direction indicated with 
the second  arrow, ``2'' we have drawn at the ``crossroad'' A in Fig. \ref{tub}, we 
observe  that the succession of colors starting from red, at constant $u$, is  
red $\rightarrow$  orange $\rightarrow$  yellow  $\rightarrow$ green 
$\rightarrow$ light blue $\rightarrow$ blue  at the intersection B.
In all cases we observe at the intersections of $S_4$ in the space $(x,\dot x,z)$ two different tubes,
coming from different directions and  having different colors. For example at the intersections  A red meets blue and at the intersection B blue meets orange. Between two 
intersections of $S_4$ we observe a smooth color variation and in all 
cases at the intersections meet different colors. This 
means that the points have different values at the 4th dimension and the 
intersections  are not intersections in the 4D space but only in the 3D 
projections.

\subsection{Rotation Numbers} 
A point that we want to investigate, is whether some properties of the rotational and tube tori are reflected in quantities, that could be defined in correspondence with the rotation numbers on the usual invariant curves around stable periodic orbits of 2D systems.
Thus we have first to define a rotation number
in the 2D and 3D projections of a 4D torus. If our 4D space of section is 
$(x,z,\dot{x},\dot{z})$, with the p.o. at ($x_0$,$z_0$,$\dot{x}_0$,$\dot{z}_0$) we 
can consider e.g. its 2D projection in (x,\.{x}) and its 3D projection in 
$(x,\dot{x},z)$. Our motivation for this is the observation, that successive 
consequents form an invariant torus by filling its surface in different ways.
Following the formation of an invariant torus on a screen as the number of 
consequents increases, one gets the impression that wires are wrapped around
the surface of the torus, following different patterns for different 
tori. This seems to be a straightforward counterpart of the different angles
at which an invariant curve is filled by successive consequents in 2D
systems, which is described by the ``rotation number'' of an invariant curve.
We want to attribute a similar number to each invariant torus. 
Let us assume  that the projection of the torus on the $(x,\dot{x})$ plane gives an 
``invariant curve'', in  the same way that the S$_i$ tori appear projected in 
Fig.~\ref{inc1}. In such  a diagram the rotation number ($rot$) is defined as  
for a usual  invariant curve around a stable periodic orbit at  
($x_0$,$\dot{x_0}$) on a 2D  surface of section. In fact ``$rot$'' is the average  
rotation angle along the  invariant curve. This quantity is different for 
different invariant curves  and  the variation of the rotation number as a 
function of the distance in the  direction of a coordinate, e.g. the 
$x$-coordinate, defines a  ``rotation'' curve (Contopoulos 2002, pg. 139). 

Accordingly, for the calculation of $rot$ for the torus in the $(x,z,\dot{x})$ 
projection,
\begin{enumerate}
\item we define an  initial  rotation  angle $\hat{r}_0$  for the initial point of an orbit  on the invariant  torus  with coordinates 
$(x_1, \dot x_1, z_1)$ as the angle formed by the vector joining this point with the central periodic orbit
($x_0$,$\dot{x}_0$,$z_0$) with the plane $(x,\dot{x})$.
More precisely:

 \begin{equation}
\hat{r}_0 = \arctan{\frac {z_{1} - z_{0}}{\sqrt{(x_{1} - x_{0})^2 + (\dot x_{1} -
    \dot x_{0})^2}}}
\end{equation}   

\item   We compute the rotation angle $\hat{r_i}$\\ between two successive 
consequents. The \textit{i-th} consequent with coordinates $(x{_i},
\dot x_{i}, z_{i})$ and the 
\textit{i+1-th} with  coordinates $(x_{i+1}, \dot x_{i+1}, z_{i+1})$  on an 
invariant torus. We define the vectors \textbf{X}  and  
\textbf{Y} as
$\textbf{X} = (x_{i+1} - x_{0}, \dot x_{i+1} - \dot x_{0}, z_{i+1} - z_{0})$ and
$\textbf{Y} = (x_{i} - x_{0}, \dot x_{i} - \dot x_{0}, z_{i} -
z_{0})$. respectively  
We find the  rotation  angle from the inner product of \textbf{X} and 
\textbf{Y} by means of the formula:  

\begin{equation}
\hat{r}_i = \arccos{\frac {\textbf{X} \textbf{Y}} { \mid \textbf{X} \mid \mid 
\textbf{Y} \mid}}
\end{equation}   

\item we finally compute the average of all $\hat{r}_{i}$'s to get
 $rot$ for the torus. 

\end{enumerate}

The rotation curves  along the x-direction, for the invariant curves  around  the simple unstable 
periodic orbit of the x1 family at Ej=$ -5.1574$,
with initial conditions
$(x_0$,$z_0$,$\dot{x}_0$,$\dot{z}_0)$ = $(0.189585220,0,0,0)$, is
given in Fig.~\ref{rcu1} (green curve). Notice that x1 is stable as regards deviations on the plane of symmetry, thus we can define a rotation number for orbits close to x1, starting on this plane.
On the x-axis of Fig.~\ref{rcu1} we give the distance from the periodic orbit in the x-direction.
Similar curves are calculated for the 
tori around  the stable  periodic orbit of the x1v1 family at the same energy (Ej = $-5.1574$),
which has initial conditions  ($x_0$,$z_0$,$\dot{x}_0$,$\dot{z}_0$)=
$(0.18939859,0.030236585,0,0)$. Two rotation curves for x1v1 are  given also in Figs.~\ref{rcu1}  
using the $(x,\dot{x})$ (blue line) and $(x,\dot{x},z)$ (red line) projections respectively.
For the calculation of the rotation numbers in these rotation curves we followed 
the definitions mentioned above. We observe that in  all  cases we have a 
similar variation of the rotation numbers. The behavior of rotation curves 
for the invariant tori around the  stable periodic orbit of the x1v1 family
has similar behavior with  the rotation curve  for the invariant curves  around 
the  simple unstable periodic orbit of the x1 family. We realize that the rotation numbers of the tube tori, follow the rotation curve and occupy the expected position in this diagram without any kind of exceptional behavior.

\begin{figure}[h]
  \begin{center}
    \hspace{-16mm}
    \resizebox{98mm}{!}{\includegraphics{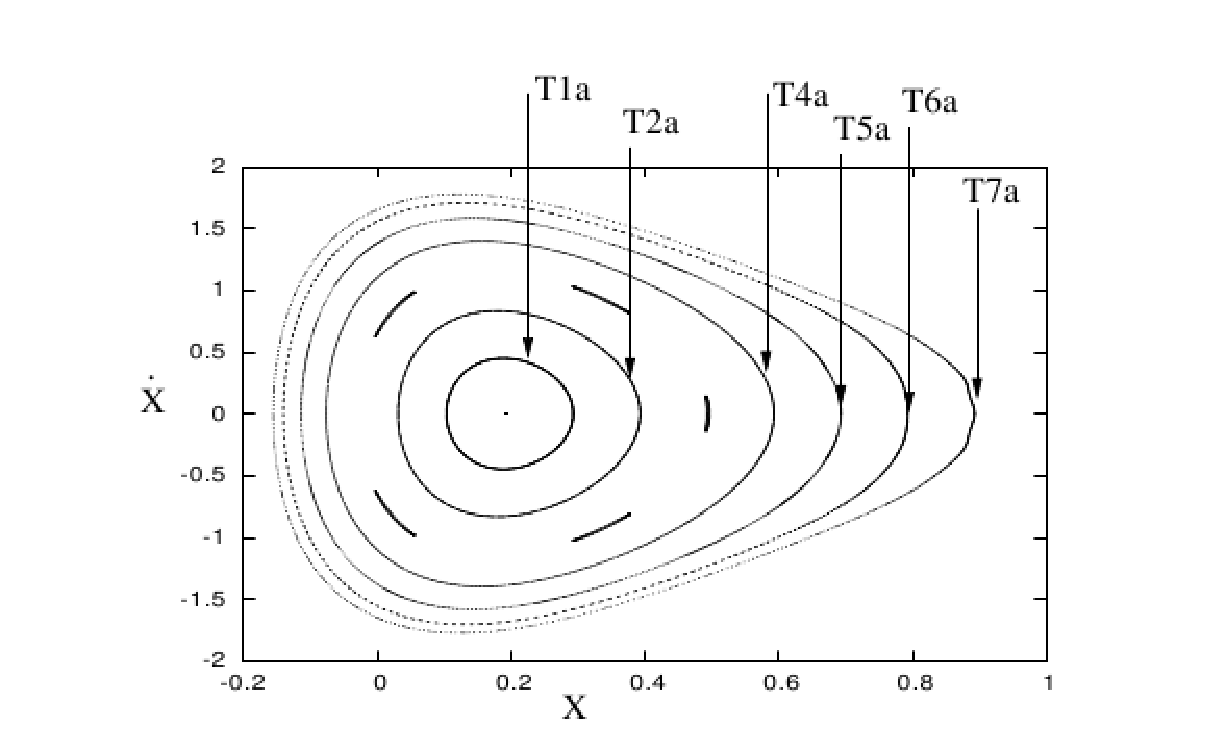}}
    \caption{The invariant curves $T_{1a}, T_{2a}, T_{4a},
      T_{5a}, T_{6a}$ and $T_{7a}$ in the $(x,\dot x)$  
      surface of section for Ej=$-5.131377$.}
    \label{inc2}
  \end{center}
\end{figure}

\begin{figure}[h]
  \begin{center}
    \resizebox{88mm}{!}{\includegraphics{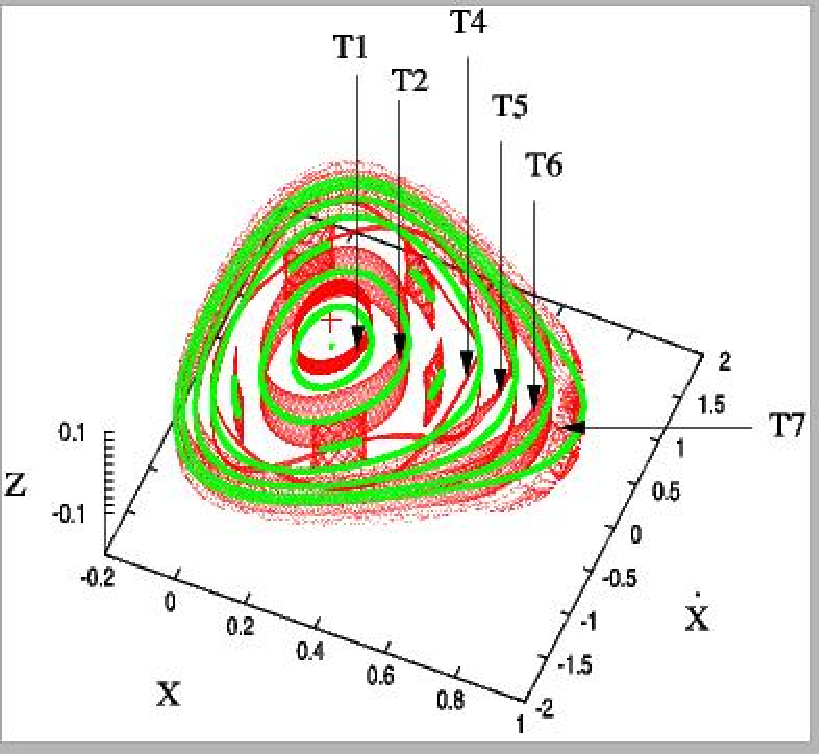}}
    \caption{The 3D $(x,z,\dot x)$ projection of the 4D surface of section for 
      Ej=$-5.131377$. Our point of view in spherical  coordinates is
      $(\theta, \phi) = (18^{o},23^{o})$. The invariant tori 
      around the initial condition of the stable p.o. x1v1 are colored red 
      and  the invariant  curves around the simple unstable p.o. of x1 on the $z=0$ plane
      are colored green. The initial conditions of $x1$ 
      and $x1v1$ are given with green and red crosses respectively. }
    \label{int3}
  \end{center}
\end{figure}

\subsection{Energy variation}

Coming back to the evolution of the stability of the families x1 and
x1v1, as described in Fig.\ref{stab1}, we study the changes introduced in the 
structure of the phase space as the energy increases. At the bifurcating point 
the initial conditions of the families x1 (parent family) and x1v1 
(bifurcating family) are identical. The x1v1 family comes in the system with two representatives at each energy (Skokos et al. 2002a). As Ej increases, we have one branch with $z>0$ and the other one symmetric with $z<0$. Let us consider here the $z>0$ case. Close to the bifurcating point, the $x_0$ 
initial condition of the stable x1v1 orbit remain  close to the  $z=0$ plane. Around the fixed point we find tori, which are practically symmetric with respect to the $z=0$ plane.   
Away from the bifurcating point, for larger values of the energy, the 
value $z_0$ in the initial  conditions of of x1v1 
increases. Thus, beyond a given energy, 
in the $(x,\dot{x},z)$ projection, the  $x_0$ initial value of the simple 
unstable  periodic orbit x1 is expected to be away from the tori  
around the stable periodic orbit x1v1, since these tori surround the x1v1 initial condition away from the z=0 plane. Here we study as an example the case for  
Ej= $-$5.131377, where  the initial conditions of the simple unstable  
periodic  orbit x1 are $(x_0,\dot x_0,z_0,\dot z_0) = (0.19317510,0,0,0)$, while 
these of  x1v1  are  $(x_0,\dot x_0,z_0,\dot z_0) = (0.19221178,0,0.06769306,0)$. 
Firstly we  explore  the phase  space around the  periodic orbit x1,
which, also at this energy, is simple unstable. We do this by 
increasing  the  $x_0$ initial condition  by 
$\Delta x_0=0.1,0.2,0.3,...,0.7$  and considering   the  $(x,\dot{x})$  surface  
of section (Fig.~\ref{inc2}). The new element in Fig.~\ref{inc2} with  respect  to the $(x,\dot{x})$ 
phase space  structure at $E_j =-5.1574$ (Fig.~\ref{inc1}) is the appearance  
of a chain of 5 islands of stability at the location  of the third 
invariant curve. These islands  correspond to the orbit with  initial 
conditions $(x_0+0.3,0,0,0)$. Thus, one  of the seven  invariant  curves of Fig.~\ref{inc1} has  been 
broken as a result of the increase of Ej.  We name  the  rest  of the 
invariant curves $T_{1a},T_{2a},T_{4a},T_{5a},T_{6a},T_{7a}$, with $T_{1a}$  
being   the closest  to  the initial conditions  of x1. 

\begin{figure}[h]
  \begin{center}
    \resizebox{90mm}{!}{\includegraphics{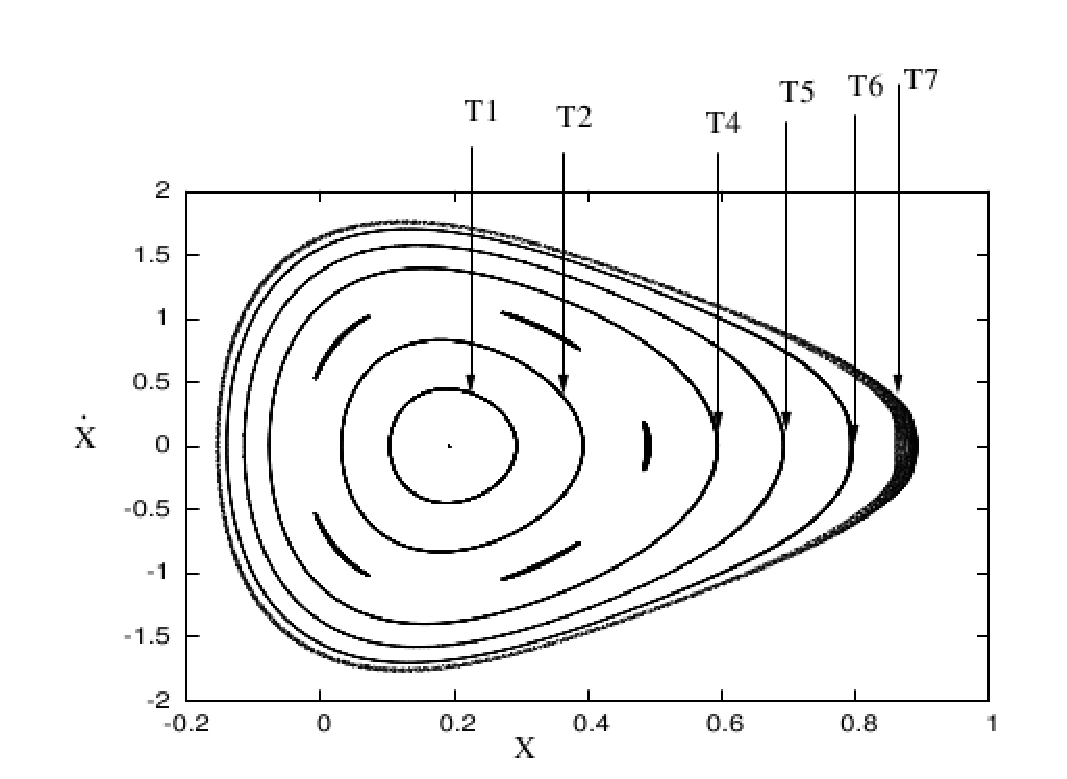}}
    \caption{The $(x, \dot x)$ cross-section space at the  neighborhood  of 
      the 3D family x1v1 for Ej=$-5.131377$.}
    \label{2D}
  \end{center}
\end{figure}

\begin{figure}
\begin{center}
\begin{tabular}{cc}
\resizebox{95mm} {!}{\includegraphics{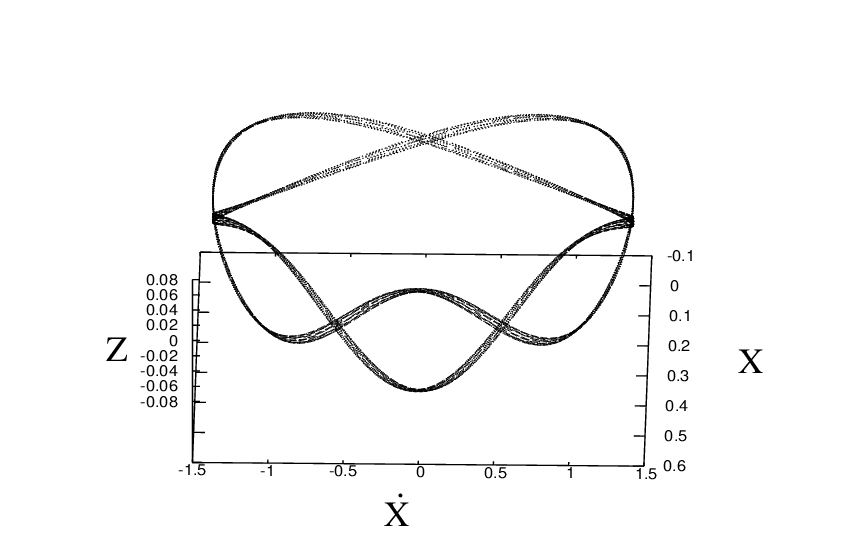}}\\
\end{tabular}
\caption{The $(x,\dot x,z)$ 3D projection of the torus $T_4$ from the point view  
$(\theta, \phi) = ( 41^{o},91^{o})$ we consider $10^5$ consequents.}
\label{proj1a}
\end{center}
\end{figure}

\begin{figure}
\begin{center}
\begin{tabular}{cc}
\resizebox{90mm} {!}{\includegraphics{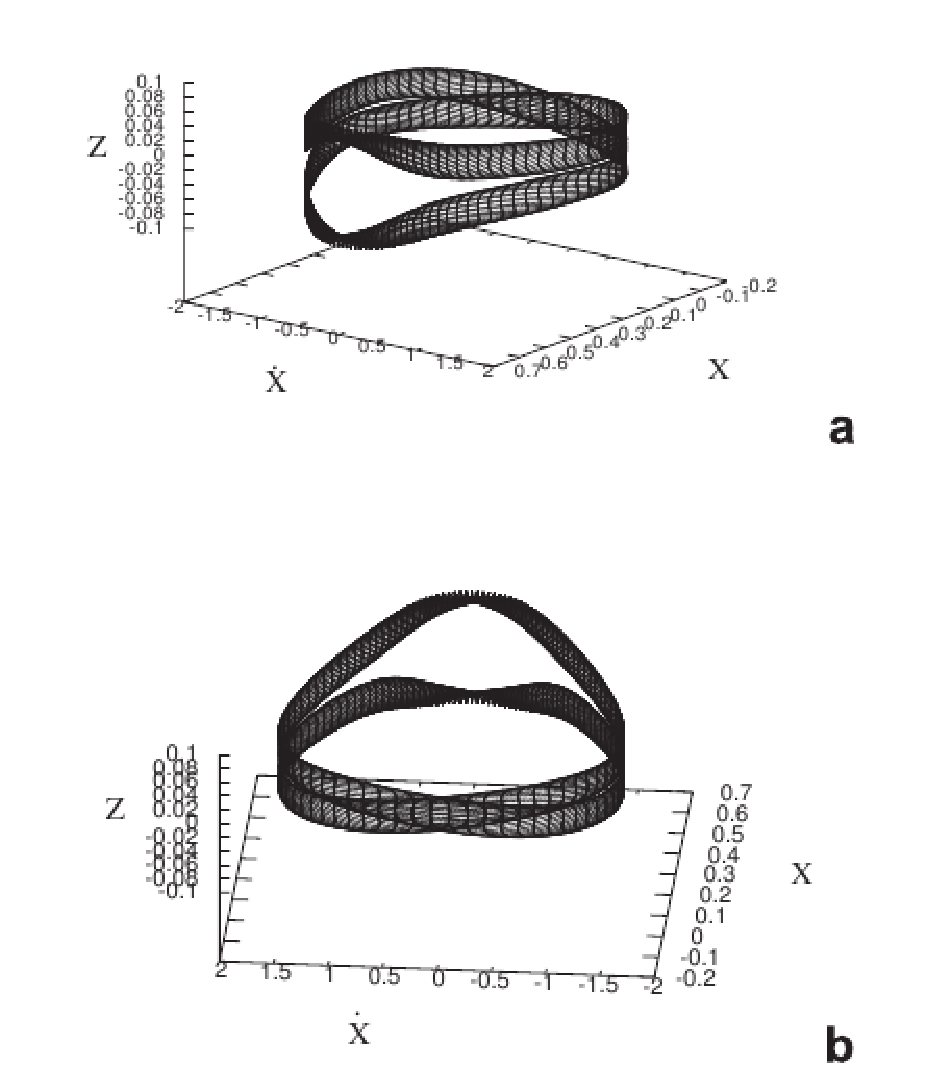}}\\
\end{tabular}
\caption{3D projections of torus $T_5$ from the point view 
$(\theta, \phi) = ( 64^{o},127^{o})$ (a) and $(\theta, \phi) = ( 48^{o}, 271^{o})$ (b).}
\label{proj2}
\end{center}
\end{figure}

As in the previous case we investigate the neighborhood of the phase space around the 3D stable  
x1v1 by  applying  the same perturbations. That means  we increase the 
$x_0$  initial  condition of x1v1  by $\Delta x_0=0.1,0.2,\dots
0.7$  and we  consider  the distribution  of the consequents  in the 
$(x,\dot{x},z)$ projection of the 4D space after $10^4$ intersections with the 
y=0 plane. We can observe the result in Fig.~\ref{int3}. The red-colored 
invariant tori around the initial condition of the stable p.o. x1v1 (red
cross at ($0.19221178,0,0.06769306)$) correspond to the seven
perturbed initial conditions we study. As in the case of the perturbed orbits 
around x1, also for x1v1  the third set of initial conditions, i.e. the one 
with $\Delta x_0=0.3$, exhibits a conspicuously different dynamical behavior. It 
forms a set of  five  small invariant tori, in correspondence to the five 
islands  of stability formed on the $(x,\dot{x})$ plane in the case of x1. Each one of the five small invariant tori has a rotational torus structure.
We name  the  remaining  six  tori that surround  the periodic orbit x1v1  
as $T_1,T_2,T_4,T_5,T_6,T_7$. The missing $T_3$  has  been substituted  by 
the five small invariant tori. In Fig.~\ref{int3} are given  also  the 
invariant  curves around the p.o. of  x1, colored  green. They lie on the 
$(x,\dot{x})$ plane. We can  observe  that the tori are roughly projected on the 
$(x,\dot{x})$ plane in the region  occupied  by the  green invariant curves. 
The projection of the 4D  x1v1  surface of section to the $(x,\dot{x})$ plane is  
given in Fig.\ref{2D}. We observe that the 2D projection of the tori 
resembles  the morphology of the  invariant curves.
Another interesting  feature of the $T_i$ invariant tori at the energy Ej=
$-5.131377$ we  study now, is that the $T_4$  torus to  the right of the set  
of the 5 small ones, as well as the tori  $T_5$ and  $T_7$ are  of different  
morphology than the  rest of the $T_i$'s. 

In particular the torus $T_4$ (Fig.~\ref{int3}) has a thin, ribbon-like structure.  
Fig.~\ref{proj1a} presents $T_4$ from a different point of view ($(\theta, \phi) = (41^{o},91^{o})$) in the $(x,\dot{x},z)$ projection of the surface
of section and  helps us understand that $T_4$ indeed intersects itself in this 
projection at five places. The behavior of the $T_4$ is similar with the 
behavior of the $S_4$ torus at the energy  Ej=$-$5.1574 (Fig.~\ref{projt}).

$T_5$  has  also a thin, 
complicated, ribbon-like structure (Fig.~\ref{proj2}). This time the torus 
intersects itself only at one region in the $(x,\dot x,z)$ 3D projection  as 
we can see  in  Fig.~\ref{proj2}a and  in  Fig. \ref{proj2}b from different
points of view of the $(x,\dot x,z)$ projection. Seven intersections in the $(x,\dot x,z)$ projection are also observed in $T_7$. $T_4,T_5$ and $T_7$ are objects that satisfy the definition of the tube tori given by 
Vrahatis et al. (1997).

 
\begin{figure*}[t]
\begin{center}
\begin{tabular}{cc}
\resizebox{95mm} {!}{\includegraphics{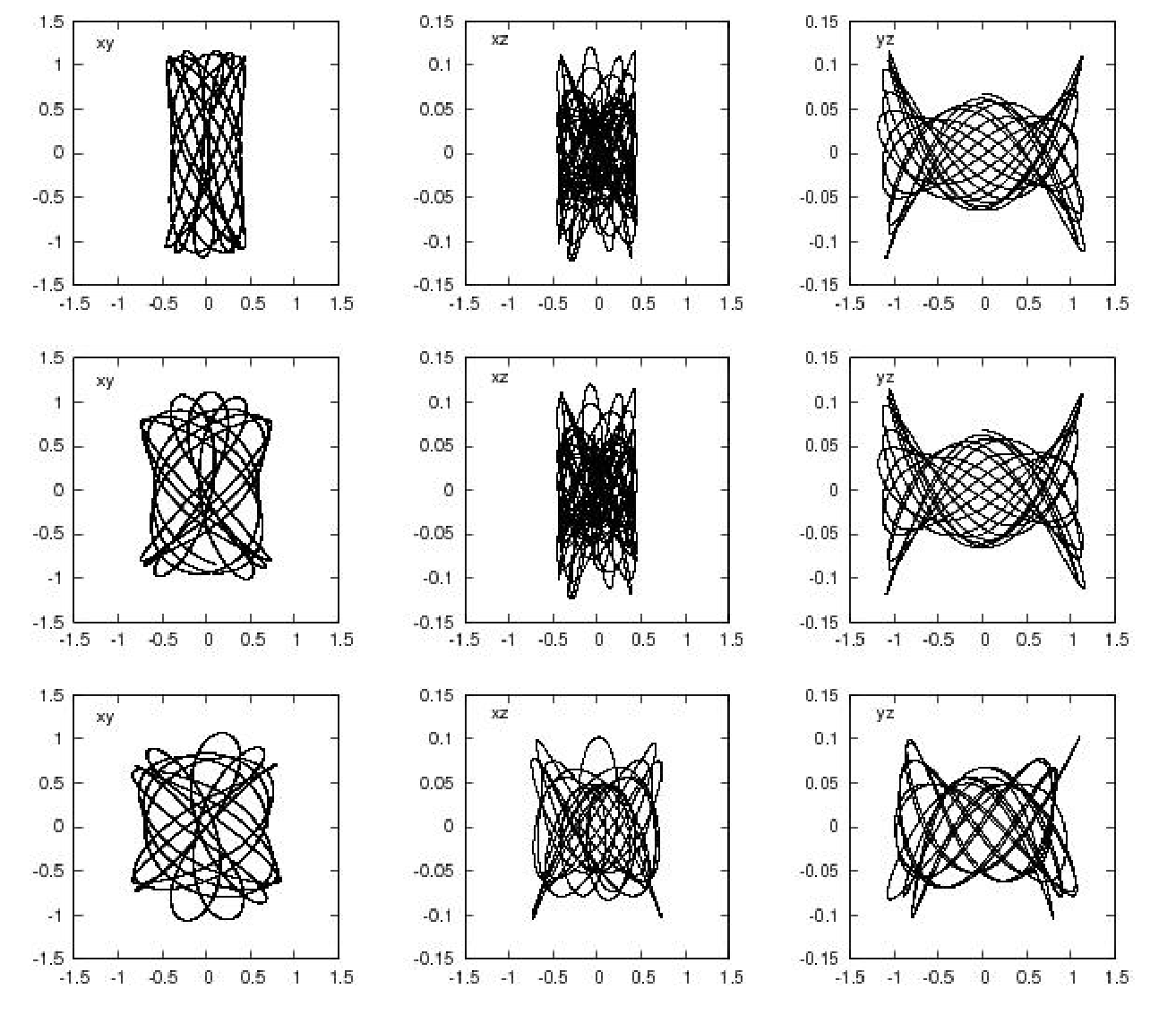}}\\
\end{tabular}
\caption{The orbits in the configuration space corresponding to the tori $T_2$ 
(first row), $T_4$ (second row) and $T_5$ (third row). Projections are
  indicated at the upper left corner of each panel. In the $(x,z)$ and $(y,z)$ projections the scales on the axes are not equal, so that we can see the detailed of the orbits.}
\label{orbt}
\end{center}
\end{figure*}

\begin{figure*}[t]
  \begin{center}
    \resizebox{100mm}{!}{\includegraphics{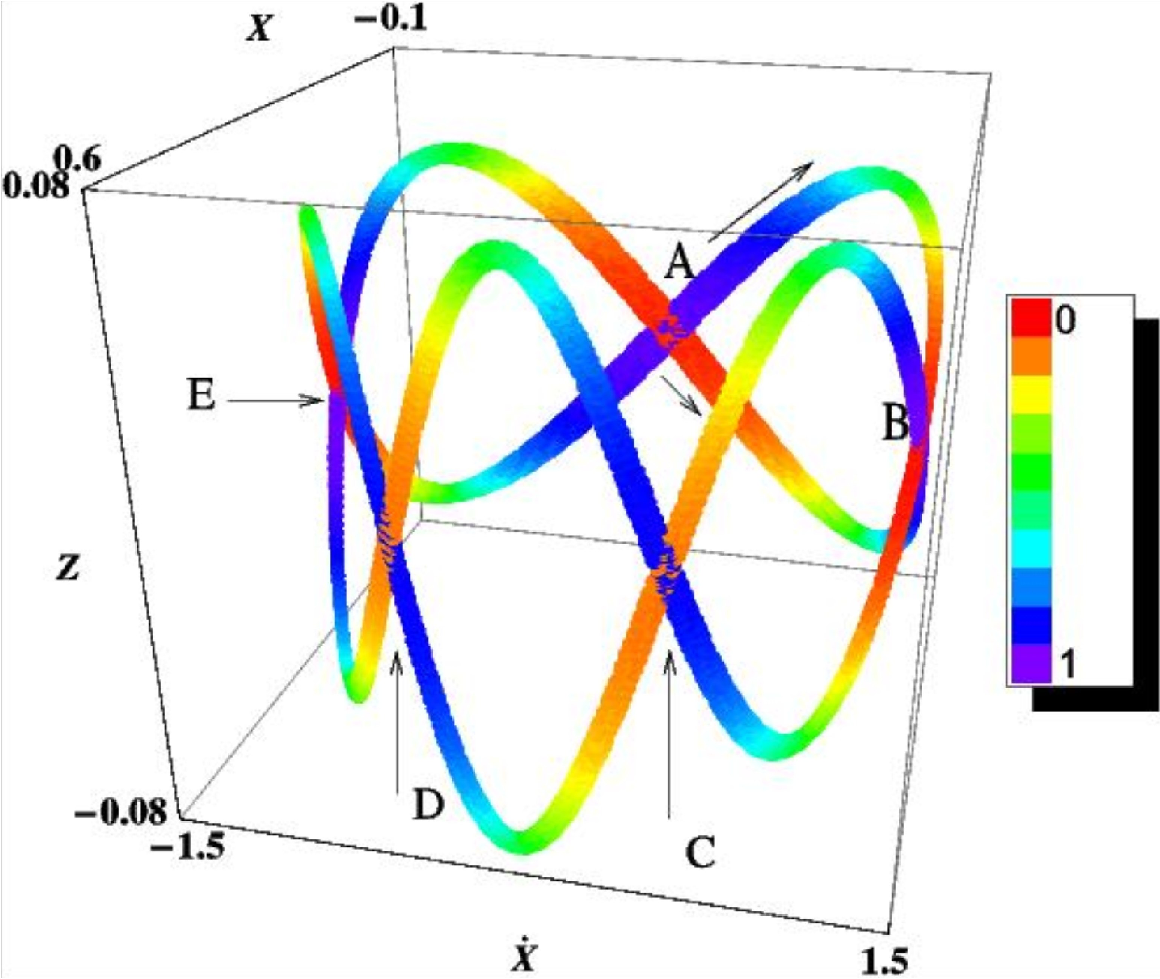}}
    \caption{ The $(x,\dot x,z,\dot z)$ 4D surface of section of the torus $T_4$ 
      for Ej=$-5.131377$. The consequents are colored according to their value
      in  the $\dot z$ coordinate. Our view angles are 
       $(\theta, \phi) = (30^{o}, 120^{o})$. The five intersections points of
      the tube torus by itself in the $(x,\dot x,z)$ are labeled with A,B,C,D
      and E.}
    \label{tub1}
  \end{center}
\end{figure*}

\begin{figure*}
  \begin{center}
    \resizebox{110mm}{!}{\includegraphics{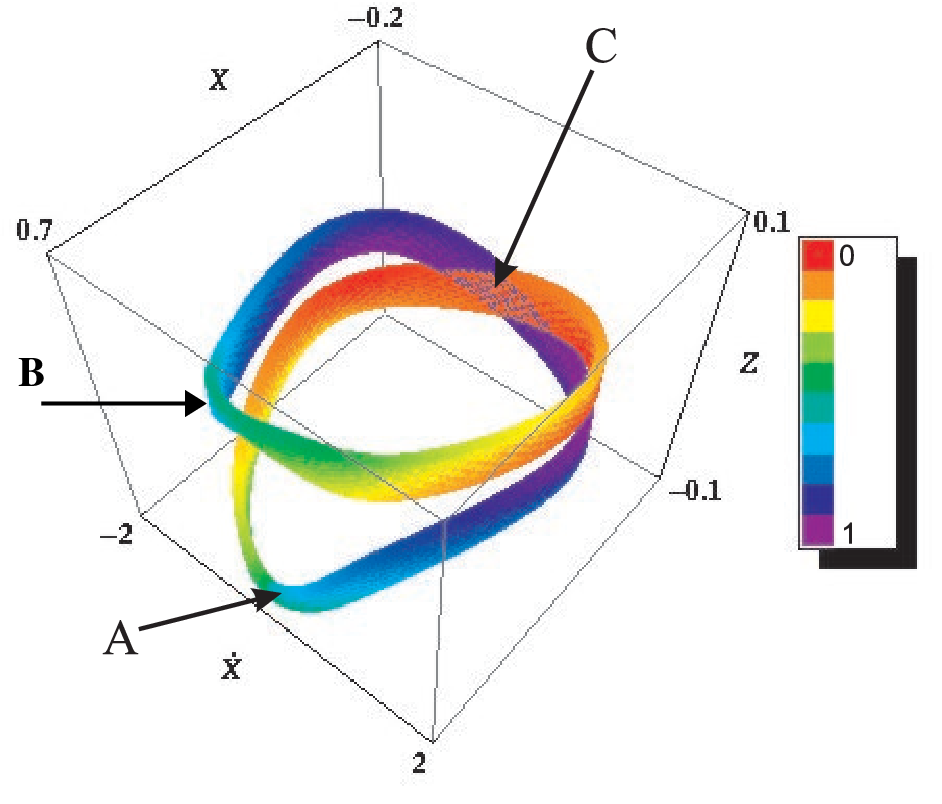}}
    \caption{ The $(x,\dot x,z,\dot z)$ 4D surface of section of torus $T_5$ 
      for Ej=$-5.131377$. The consequents are colored according to their value 
      in  the $\dot z$ coordinate.  Our view angles are 
      $(\theta, \phi) = (30^{o}, 72^{o})$. }
    \label{rott2b}
  \end{center}
\end{figure*}

As we said, in the $(x,\dot x)$ projection the differences in the 
morphology of $T_4,T_5$ and $T_7$ from the rest $T_i$'s cannot be seen
(Fig.~\ref{2D}). Trying to understand if the difference we observe in the morphologies of the 3D projections of the tori reflect some morphological differences of the orbits in the configuration space we compare the orbits corresponding to the $T_2$ (rotational torus), with those of the tori $T_4$ and $T_5$ (tube tori) in Fig.~\ref{orbt}.
Their morphology is the expected for quasiperiodic orbits trapped close to a stable periodic orbit, in our case x1v1. 
The $(x,y)$ projections of $T_2, T_4$ and $T_5$ are practically identical with the $T_{2a}$,$T_{4a}$ and $T_{5a}$ orbits around x1 for the same Ej. There is no obvious morphological feature that distinguishes the three 3D orbits among themselves also in the $(x,z)$ and $(y,z)$ projections.
 
$T_4$ offers the opportunity to study in detail the structure of one more tube torus in the 4D space and compare it with the one depicted in Fig.~\ref{tub}, which we called $S_4$.
Applying the color-rotation method  also in
this case, we observe,
that the structure of $T_4$ (Fig.~\ref{tub1}) is similar with that of $S_4$ (Fig.~\ref{tub}).
We observe in Fig.~\ref{tub1}, that moving along the tube 
from A towards B, following the directions indicated with arrows, we follow
the succession of the colors of the color bar from one side to the other. At
the intersection regions, in the $(x,\dot x,z)$ projection, the red color meets
 blue. This  means  that  these  regions  are not intersections in the 
4D space. Again here the intersections appear only at the 3D projections.

Different colors at the intersections characterize all tube tori we studied with the color-rotation method. A final example is given for the case of $T_5$. As we said before the torus $T_5$ is very thin, has almost a
ribbon-like structure, and intersects itself at one region (Fig.~\ref{proj2}).
In Fig.~\ref{rott2b}   
we depict a projection in the 3D subspace $(x,\dot x,z)$ and we color it according to 
the values of the consequents in the 4th dimension $\dot z$. By
moving counterclockwise from the region A we observe again the smooth color variation. The 
colors change from green to light blue and then to blue until we 
reach the region C. At the region C $T_5$  intersects  itself in the projection
$(x,\dot x,z)$. Then, moving always counterclockwise, the  succession of the 
colors  continues as blue $\rightarrow$ light blue  $\rightarrow$  green  
at the region B and finally comes back to the region C. At the region C we observe that blue color
meets red. This means that the points have different values 
in the 4th dimension and  C is not a region that we have a  real intersection 
of $T_5$ in the 4D space. If we will continue our counterclockwise journey along the thin tubes of $T_5$ we will reach A following always a smooth color variation. Of special interest  in this case is the folding of the thin, ribbon-like torus at the regions A and B. At these two regions the internal  surface of the torus becomes
external  surface and vice-versa. This warping is encountered along the tubes of several other tori we studied. $T_7$ intersects itself in the 3D projections at 7 regions, while the tubes warp in the journey from one intersection to the other.

\subsubsection{Rotation numbers}
The rotation curves in the x-direction along the invariant curves around x1 and along the projected tori around x1v1 in the $(x,\dot{x})$ section, as well as the rotation curve from the tori around x1v1 in the $(x,\dot{x},z)$ projection, calculated using the definition of Eq.(7), are very similar as in the case we presented above at  Ej = $-5.1574$. Also here, at E$_J = -5.131377$, there is no special variation of the rotation curve at the locations of the tube tori. The variation of the rotation curves follows the general rules of these curves (see Contopoulos 2002). The main difference is that at the location of the stability islands around x1 and the
tori-chains around x1v1 we find tiny plateaus on the rotation curve as expected.
 
\begin{figure*}[t]
  \begin{center}
    \resizebox{80mm}{!}{\includegraphics{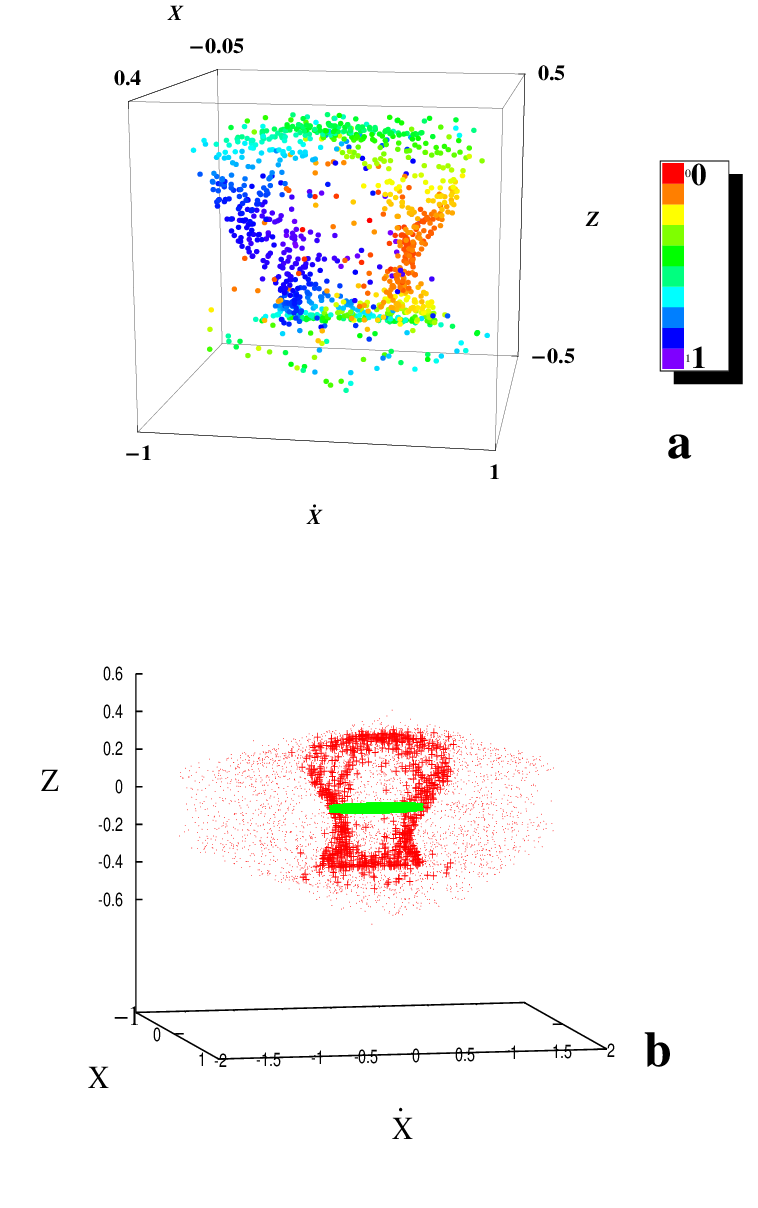}}
    \caption{The 3D $(x,z,\dot x)$ projection of the 4D surface of section in the neighborhood of x1 for 
      Ej=$-5.207$. The initial conditions of the orbit  are $(x_0,\dot x_0,z_0,\dot z_0)$ $=(0.28312784,0,0.41,0)$ (see text). (a) the first 1100 consequents colored according to their position in the 4th dimension $\dot z$. The point of view is $(\theta, \phi) = (78^{o},82^{o})$. (b) The first 1.2 $\times 10^4$ consequents (red dots). The initial condition of x1 is indicated with a green dot, while the green invariant curve around it correspond to the orbit with initial conditions $(0.28312784,0,0,0)$. The red points diffuse in the phase space after staying on a toroidal surface for about 900 consequents.}
    \label{Px5}
  \end{center}
\end{figure*}

\begin{figure*}
  \begin{center}
    \resizebox{80mm}{!}{\includegraphics{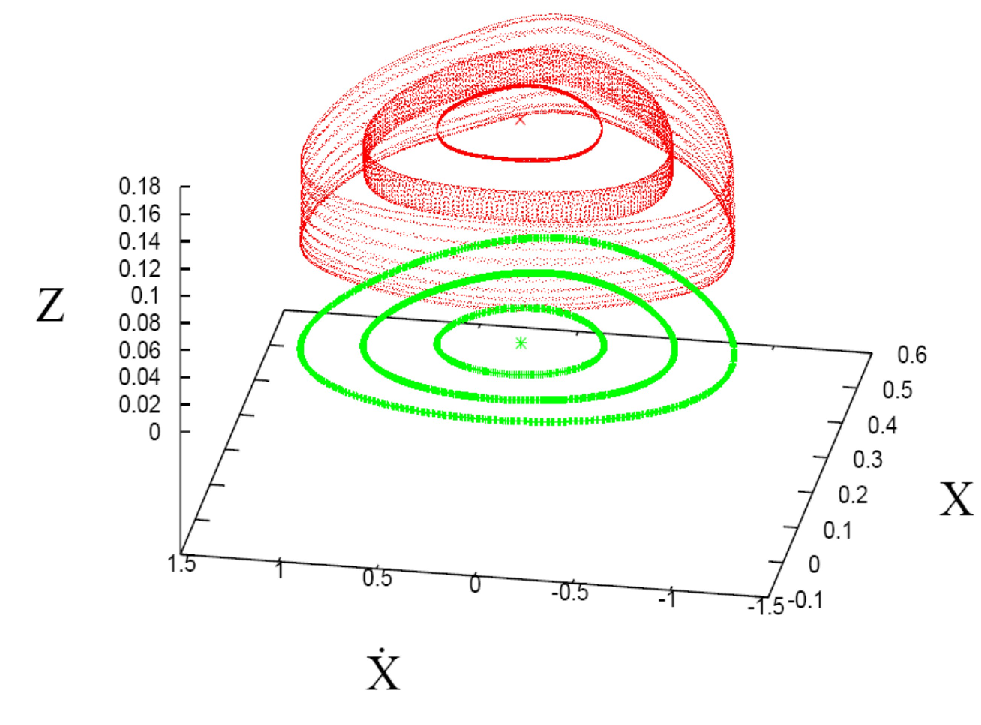}}
    \caption{The 3D $(x,z,\dot x)$ projection of the 4D surface of section at
      Ej=$-4.98996$ for three invariant tori around $x1v1$ and three invariant curves around x1 (green lines). Our point of view is
      $(\theta, \phi) = (56^{o},280^{o})$. The initial condition of $x1$ is indicated with a green star and
      the initial condition of $x1v1$ is given with a red ``$\times$''.}
    \label{Pz1}
  \end{center}
\end{figure*}

\begin{figure*}[t]
  \begin{center}
   \resizebox{140mm}{!}{\includegraphics{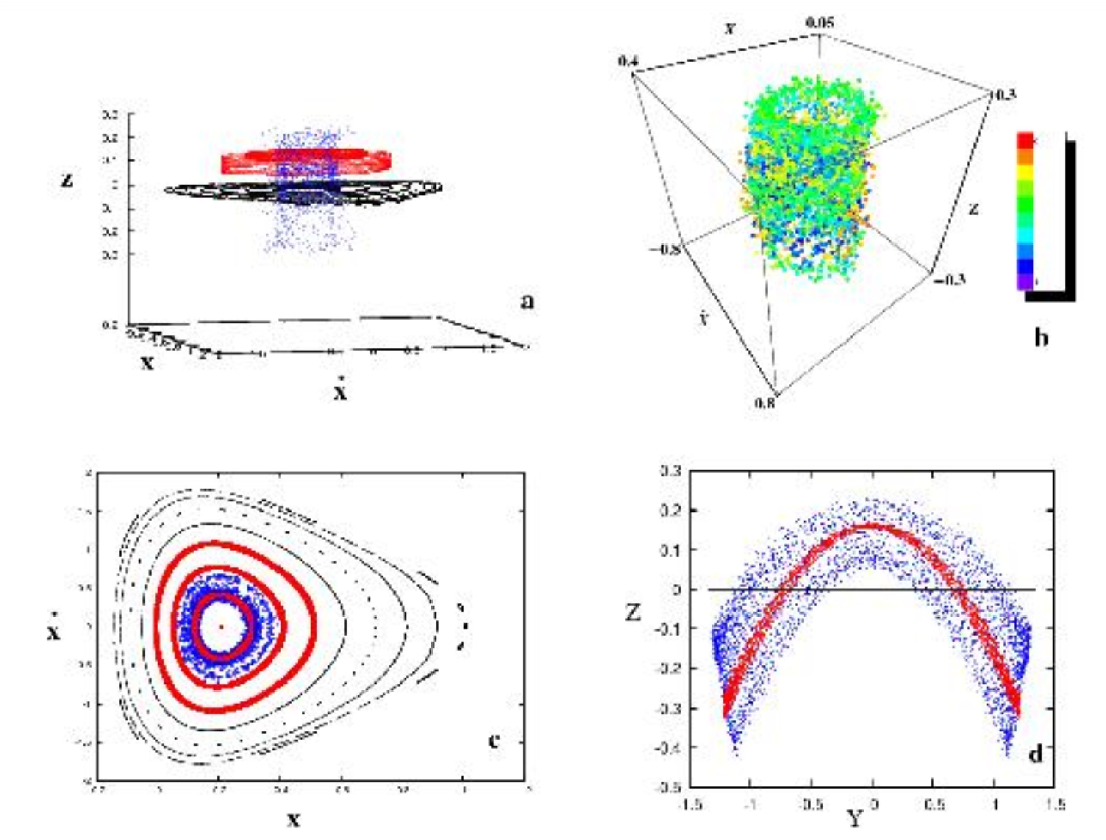}}
    \caption{(a) A chaotic orbit at Ej=$-4.98996$ in the 3D $(x,z,\dot x)$ projection (point of view 
      $(\theta, \phi) = (55^{o},263^{o})$) of the 4D surface of section is given with blue dots. We also give three rotational tori around x1v1 (red), and seven invariant curves around x1 in the $(x,\dot{x})$ plane. (b) The 4D surface of section of the cloud of blue points we give in (a) (point of view $(\theta, \phi) = (60^{o},45^{o})$). It is characterized by scattering of their colors in the 4th dimension. (c) The $(x,\dot{x})$ surface of section for the orbits in (a). We see that the blue points extend inside as well as outside the innermost rotational torus. (d) The chaotic orbit (blue points) together with the quasi-periodic x1v1 orbit corresponding to the innermost rotational torus in (a).}
    \label{Pz3}
  \end{center}
\end{figure*}

\begin{figure*}[t]
  \begin{center}
    \resizebox{100mm}{!}{\includegraphics{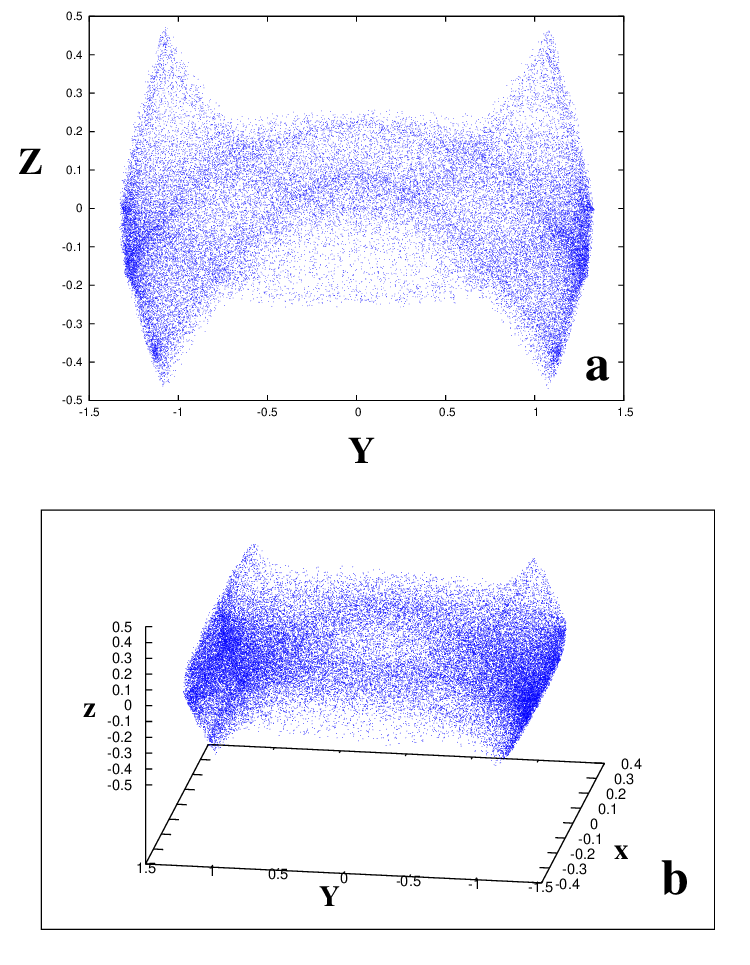}}
    \caption{ (a) The $(y,z)$ profile of the orbit corresponding to the cloud of blue points in Fig.~\ref{Pz3}a. (b) The same orbit in the configuration space from a point of view $(\theta, \phi) = (63^{o},280^{o})$.}
    \label{bo1000}
  \end{center}
\end{figure*}

\begin{figure*}
 \begin{center}
  \resizebox{120mm}{!}{\includegraphics{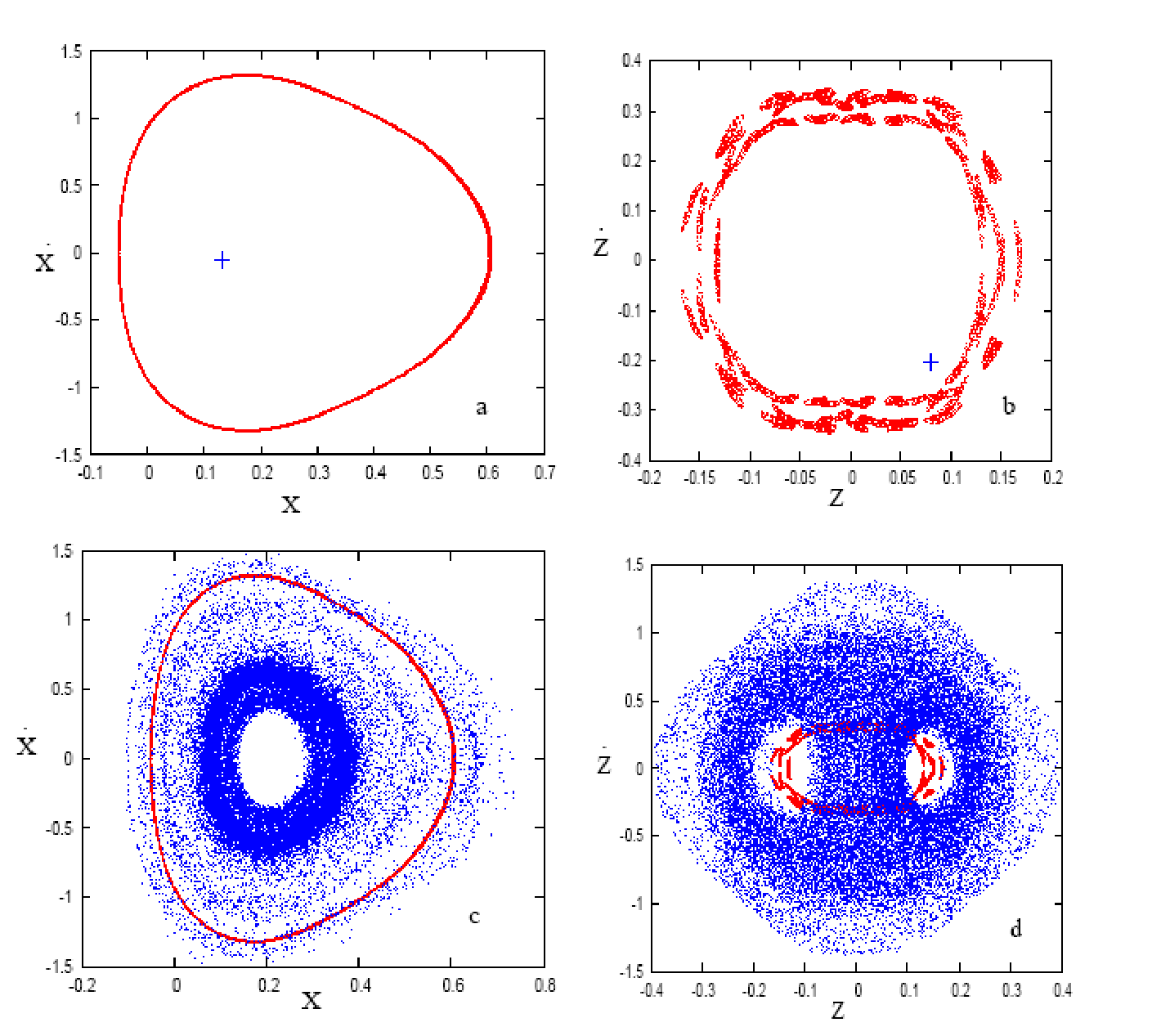}}
   \caption{The relative location of the consequents of the chaotic orbit of Fig.~\ref{bo1000}, with respect to the location of the tube torus ``X4'' (see text). In (a) and (b) we give with a blue cross, the location of a single consequent that it is projected in the interior of the $(x,\dot{x})$ and $(z,\dot{z})$ projections of ``X4'' respectively. In (c) and (d) we give all the $1.2 \times 10^4$ consequents of the orbit, clearly showing that we have points inside and outside ``X4''.}
   \label{arn}
 \end{center}
\end{figure*}

\section{Perturbations in the z- and \. z-direction}
The radial perturbations we studied until now, gave us an understanding of the
structure of the tori in the neighborhood of stable periodic orbits in our
galactic type Hamiltonian System. However, of particular interest is the
evolution of the tori also under vertical perturbations. Especially in 3D
galactic systems we want to know how, and in what extent, stars are trapped
away from the equatorial plane. This study is related with the investigation
of morphological features as boxy and peanut-shaped bulges in disk galaxies
(Patsis et al. 2002).

\subsection{Spaces of section before the $S \rightarrow U$ transition}
Going back to the stability diagram given in Fig.~\ref{stab1}, that describes the stability evolution of the families x1 and x1v1, we study first the phase space at an energy before the $S \rightarrow U$ transition, i.e. at Ej=$-$5.207. In this case, on the $(x, \dot x)$ surface of section the only 2D simple periodic orbit we have with $x>0$ is x1 and it is stable.  Around x1, we find invariant curves as expected, if we perturb its $x_0$ initial condition. As an example we consider the invariant curve, which results from the initial conditions $(x_0 +\Delta x_0,\dot x_0,z_0,\dot z_0)$ $=(0.18312784 + \Delta x_0,0,0,0)$, with $\Delta x_0$=0.1. If we ``perturb'' this quasi-periodic orbit, i.e. if we increase further the $z_0=0$ initial condition by $\Delta z_0$=0.1, 0.2 \dots, and integrate again our initial conditions, we encounter tori around x1. These are both  rotational and tube tori. The tori, for small $\Delta z_0$ surround the invariant curve we perturbed. For $\Delta z_0 \geq $0.3 the tori appear distorted and inclined with respect to the $z=0$ plane. However, for $\Delta z_0$=0.41 the dynamical behavior at the neighborhood of the quasi-periodic orbit changes. About the first 900 consequents of this orbit form an object that resembles a rotational torus. By this we mean that initially the consequents stay approximately on a toroidal surface. However,
the following consequents diffuse in the phase space and occupy a larger volume in it. In Fig.~\ref{Px5}a we observe the $(x,\dot x,z)$ projection of the 4D surface of section for this orbit, where the 4th dimension is represented by the colors of the 4th $\dot z$ coordinate. The number of consequents in Fig.~\ref{Px5}a is 1100. They form a (rotational) toroidal surface, with a smooth color variation around it as we can see by comparing the colors on its surface with the color bar to the right of the figure. We note, that 900 consequents do not suffice to fill densely the surface of the toroidal object. Thus, on Fig.~\ref{Px5}a we observe distinct points rather than a toroidal object. However, starting e.g. from upper left, we observe the succession of colors from green $\rightarrow$ yellow $\rightarrow$ orange $\rightarrow$ red $\rightarrow$ orange $\rightarrow$ yellow $\rightarrow$ green $\rightarrow$ to blue shades. In Fig.~\ref{Px5}a some points have already started deviating from the ``torus'' and soon they occupy a larger area of the phase space. In Fig.~\ref{Px5}b, the dense cluster of red points corresponds to the 900 consequents we plotted colored in Fig.~\ref{Px5}a. The rest of the red points that occupy a larger area of the phase space belong to the same orbit, which has been integrated now for $1.2 \times 10^4$ consequents. In the same figure we give as well the invariant curve around x1 at this energy, drawn with green color. This dynamical behavior indicates stickiness (Contopoulos \& Harsoula 2008). Qualitatively, we find similar results if we increase $\dot z_0$ instead of $z_0$, in the initial conditions $(x_0,\dot x_0,z_0,\dot z_0)$ $=(0.28312784,0,0,0)$ following the procedure described above.

\subsection{Spaces of section after the $S \rightarrow U$ transition}
As we already have seen, after the $S \rightarrow U$ transition, x1 becomes simple unstable, while the 3D stable family  x1v1 has been bifurcated. The energy Ej=$-$5.131377 is already beyond the bifurcating point (Fig.~\ref{stab1}). We have examined the phase space structure in the neighborhood of x1 and x1v1 by perturbing their $x_0$ initial conditions and the result is depicted in Fig.~\ref{int3}. At energies not far away from the bifurcating point A (Fig.~\ref{stab1}), as is Ej=$-5.131377$, we could always find a ``$\pm \Delta z_0$'' perturbation of the quasi-periodic orbits, so that the initial condition $(x_0 + \Delta x_0,0,\pm \Delta z_0,0)$ gives a torus surrounding the invariant curve $(x_0 + \Delta x_0,0,0,0)$, with $(x_0,0,0,0)$ being the initial condition of x1. 

However, at larger energies this was not always possible. We could find invariant curves around x1, that deviate in the phase space from the invariant tori around x1v1 at the same energy. For example at Ej=$-$4.98996, we find invariant curves around x1 and invariant tori around x1v1 by perturbing the $x_0$ initial conditions of both families. In Fig.~\ref{Pz1} we give in the 3D $(x,z,\dot x)$ projection the invariant curves around x1 for $x_0 + \Delta x_0$ = 0.1, 0.2 and 0.3 (green curves) and the corresponding tori around x1v1 applying the same perturbation to its initial conditions. The invariant tori ``float'' above the invariant curves. We observe that as we depart from the x1v1 initial condition the invariant tori become thicker in $z$. Beyond a certain $x_0 + \Delta x_0$ perturbation there are invariant tori that intersect the $(x, \dot x)$ plane (not depicted in Fig.~\ref{Pz1}). 

In the case depicted in Fig.~\ref{Pz1} we successively ``perturb'' the initial conditions of the quasi-periodic orbits that correspond to the red invariant tori by $- \Delta z_0$ trying to find other invariant tori, that their projections in $z$ reach the $(x, \dot x)$ plane. Following this procedure we find a last invariant torus with  $(x_0 + 0.1,0,z_0 + \Delta z_0,0) = (0.3160689,0,-0.08,0)$, i.e. for $\Delta z_0 =-0.08$. For $\Delta z_0 = -0.09$ instead of a torus we find a cloud of points at the area between the invariant curves and the invariant tori. This cloud is given in Fig.~\ref{Pz3}a with 1.2$\times 10^4$ blue points, together with the three rotational tori around x1v1 depicted in Fig.~\ref{Pz1} (colored red) and seven invariant curves around x1 on the $(x,\dot{x})$ plane (black curves). For about the first 3000 consequents, the cloud has a vague toroidal structure in the $(x,\dot{x},z)$ projection.
In Fig.~\ref{Pz3}b we apply the color-rotation method in the set of blue points of Fig.~\ref{Pz3}a, which we color according to their value in the $\dot{z}$ coordinate. It becomes clear that the points do not have a smooth variation in the 4th dimension. The same color mixing is also present in the figure with fewer consequents. Particularly helpful is the $(x,\dot{x})$ projection of Fig.~\ref{Pz3}a, given in Fig.~\ref{Pz3}c. In this projection we can clearly observe, that blue points of the cloud can be found inside, as well as outside of the red innermost invariant rotational torus. The vast majority of the points diffuses outside the the $(x,\dot{x})$ projection of that torus between 4$\times 10^3$ to 5$\times 10^3$ consequents, building a second ring at the outer side of the torus of lower intensity than the ring in its inner side. It is interesting to realize that such chaotic orbits may contribute to the thickening of observed structures as the peanut-shaped bulges of disk galaxies. The backbone of such peanut-shaped structures is the x1v1 family (Patsis et al. 2002). In Fig.\ref{Pz3}d the trapped quasi-periodic orbit around the stable p.o. x1v1 at this energy (Ej=$-4.98996$) is plotted with red dots, while the blue points give the chaotic orbit that we plot in Fig.~\ref{Pz3}b in the configuration space ($(y,z)$ projection). The blue orbit in Fig.\ref{Pz3} is integrated for a time corresponding to 0.2~Hubble times, i.e. in a time generally believed that morphological features as the galactic bars can well survive (see e.g. Debattista et al. 2006).

For longer times, of the order of 0.7 of a Hubble Time, the shape of the ``blue'' orbit still follows the peanut morphology, having populated both branches of the stable p.o. x1v1 (Patsis et al. 2002) at this energy. Thus, the peanut shape is fully developed, despite the fact that the orbit reaches $z$ values 1.5 times larger than the quasi periodic orbit in the neighborhood of x1v1 (innermost rotational torus). In Fig.~\ref{bo1000}a we give the $(y,z)$ projection for comparison with the models of periodic orbits in Patsis et al. (2002), while in  Fig.~\ref{bo1000}b we give the same orbit from  the point of view ($(\theta, \phi) = ( 63^{o},280^{o})$), so that it becomes clear the relation between the peanut side-on profile of the orbit and its ``pencil-sharpener'' 3D morphology. This class of orbits may be very important for the kinematics of peanut-shaped bulges of disk galaxies, as they increase the dispersion of the velocities of the stars, participating in this structure without destroying its morphological profile. These orbits are chaotic, but for times important for Galactic Dynamics have consequents that stay close to the invariant tori of x1v1 in the phase space. However, for larger times they diffuse away from the invariant tori in the phase space.
 
None of the blue consequents presented in the $(x,\dot{x})$ projection (Fig.~\ref{Pz3}c) inside the innermost red torus could be found projected simultaneously in the interior of the torus in the $(z,\dot{z})$ projection. However, this was not true for the location of consequents relative to tori at larger distances from the periodic orbits than the three rotational tori we considered in Figs.~\ref{Pz1} and \ref{Pz3}. An example is given for the torus we obtained by perturbing by $\Delta x_0$=0.4 the x1v1 initial conditions. This is a tube torus and let us call it ``X4''. In Fig.~\ref{arn} we give with a blue cross the position of a consequent which is simultaneously projected in the interior of ``X4'' of the $(x,\dot{x})$ (Fig.~\ref{arn}a) as well as of the $(z,\dot{z})$ projection (Fig.~\ref{arn}b). It is one of the consequents depicted in Figs.~\ref{arn}c and Fig.~\ref{arn}d, where we clearly observe that we have consequents of the ``blue'' orbit inside and outside the tube torus ``X4''. This tells us that in this case takes place the phenomenon of Arnold diffusion (see e.g. Contopoulos 2002, pg. 344). Nevertheless the time scale for this exceeds one Hubble Time. Thus, in this particular case the consequences for the dynamics in the neighborhood of x1v1 are negligible. However, in a forthcoming paper we investigate in detail this as well as other cases of Arnold diffusion in the Hamiltonian system of Eq.~4 and Eq.~5. 

Before closing this section we want to underline that the same procedure can be repeated by perturbing the $\dot z_0$ initial condition, reaching the same results as regards the appearance and the foliation of the tori in the 4D space.

\section{The evolution of the tube tori as the perturbation increases}
Finally we examine the evolution of the invariant tori 
as the perturbation of our system increases, starting from the axisymmetric case 
($q_a=1$ and $q_b=1$). In our model we increase the perturbation by
increasing the triaxiality of the system through the parameters $q_a$ and 
$q_b$. At first we examine the axisymmetric  case 
and then we introduce a small perturbation ($q_a=1.01$ and $q_b=1$), in order to study possible qualitative differences that are introduced in the system by the perturbation.

\begin{figure}[t]
\begin{center}
\begin{tabular}{cc}
\resizebox{85mm}{!}{\includegraphics{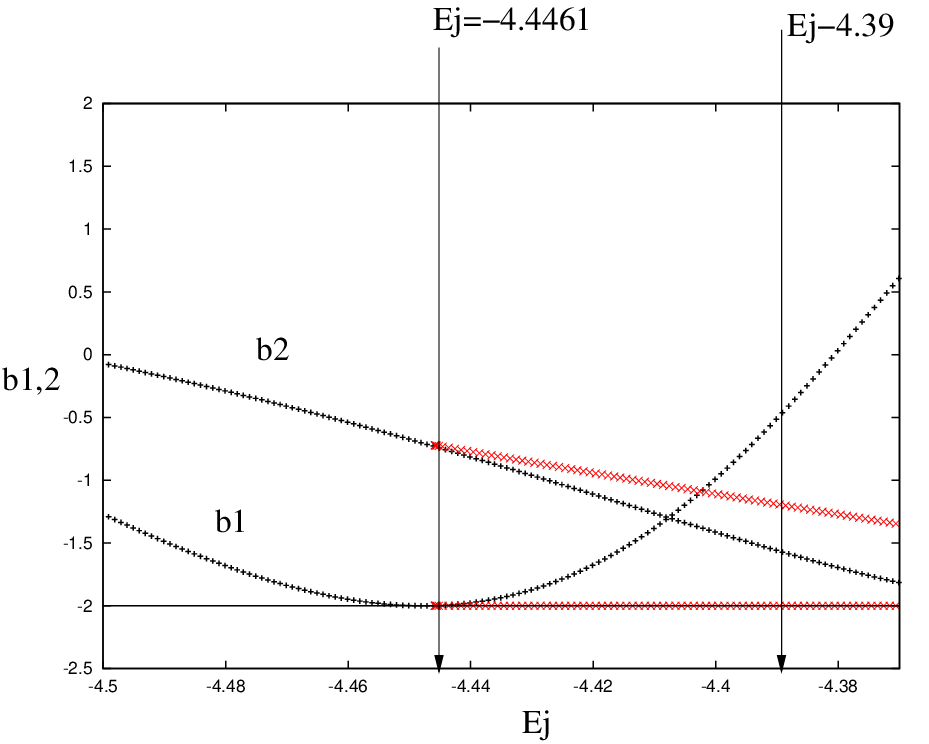}}\\
\end{tabular}
\caption{Stability Diagram for $-4.5 < $Ej $< -4.37$, that gives the stability 
indices of the family x1 and its bifurcating family of p.o. x1v1, in the
axisymmetric case of our model ($q_a=1,q_b=1)$.}
\label{stab2a}
\end{center}
\end{figure}

\begin{figure}
  \begin{center}
   \resizebox{79mm}{!}{\includegraphics{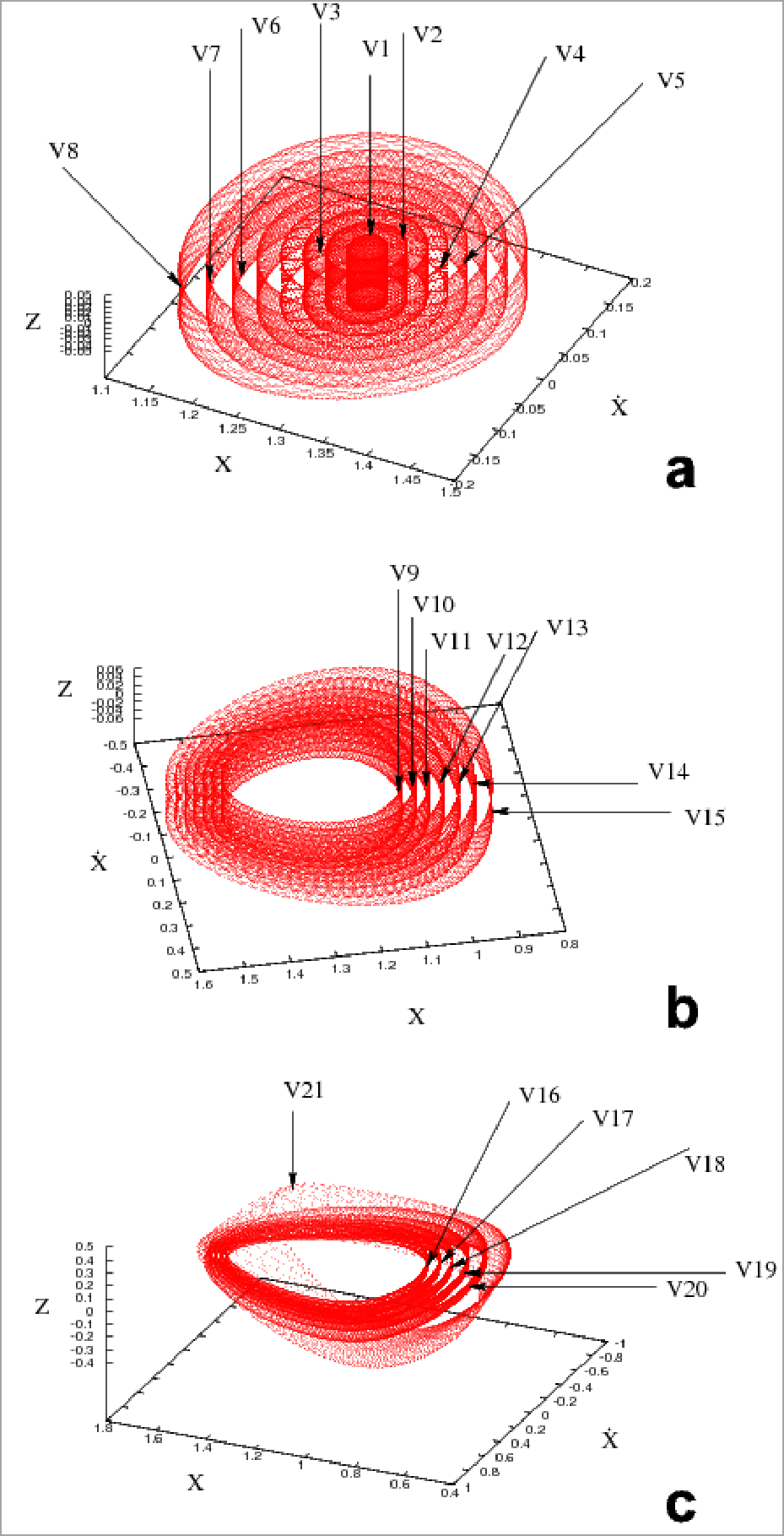}}\\
   \caption{$(x,\dot x,z)$ 3D projections of the  4D surface of  section  in 
     the axisymmetric case, for Ej=$-4.4461$, at the neighborhood of the 
     marginally stable p.o. x1v1.  (a) The first 8 invariant tori V1\dots V8 
     (point of view $(\theta, \phi) = (38^{o}, 211^{o})$). (b) The invariant 
     tori V9\dots V15, (point of view $(\theta, \phi) = (36^{o}, 244^{o})$). 
     (c) The last six tori of our sample, V16\dots V21, (point of view 
     $(\theta, \phi) = (49^{o}, 195^{o})$). We observe the deformation of the 
     last torus V21.}
    \label{proji2}
  \end{center}
\end{figure}

\begin{figure}
 \begin{center}
   \resizebox{100mm}{!}{\includegraphics{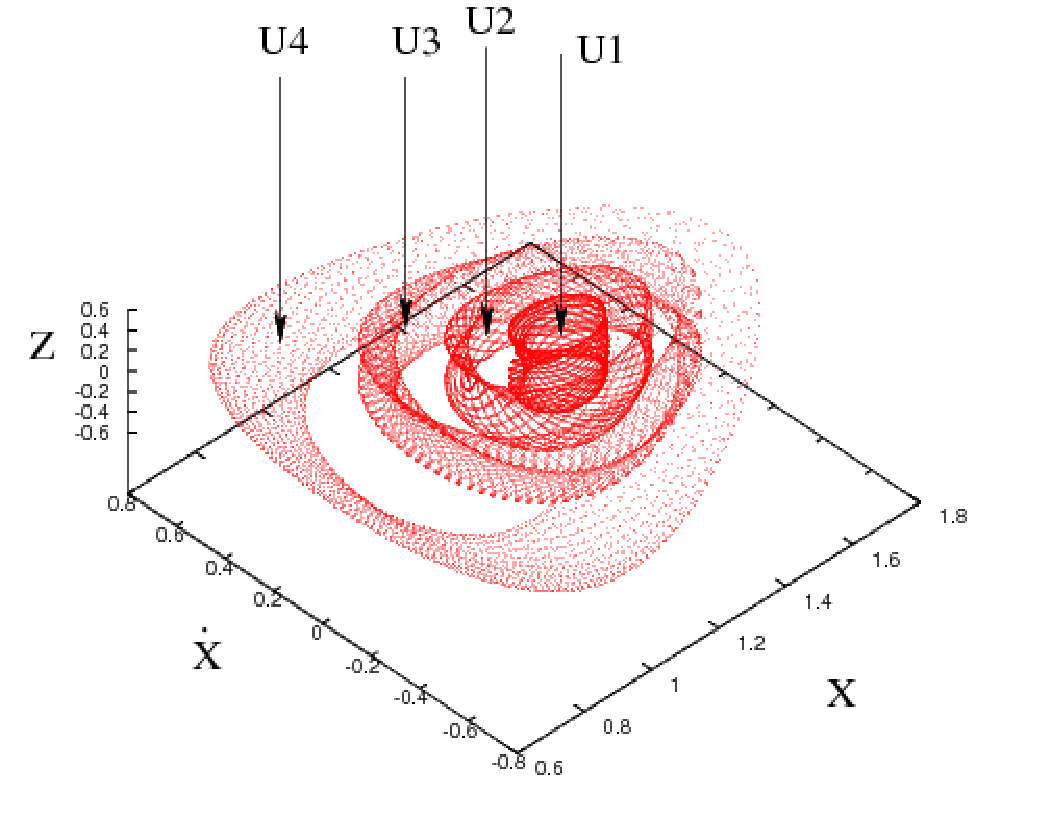}}\\
   \caption{The $(x,\dot x,z)$ 3D projection of the   
      4D surface of section in the neighborhood of the p.o. x1v1 for Ej=$-4.39$. Our point of view 
      is  $(\theta, \phi) = (27^{o}, 316^{o})$. }
   \label{proji1}
 \end{center}
\end{figure}

\subsection{The  Axisymmetric case ($q_a=1, q_b=1$)}
In 3D axisymmetric potentials the stability index of the central family, which is related with perturbations vertical to the equatorial plane, becomes tangent to the b=$-2$ axis at the vertical resonances (e.g. Patsis \& Grosb{\o}l 1996).
Fig.~\ref{stab2a} gives the evolution of the stability  of the central family 
of periodic orbits of our system, x1, and its vertical bifurcation (x1v1) 
for $-4.5 < $Ej$ < -4.37$. We observe that x1 (b1, b2 indices with black lines)  
is  initially stable and at Ej=$-$4.4472 we have a tangency of b1 with the
$-2$ axis. This tangency corresponds to the vertical 2/1 resonance. At this point it is bifurcated a new family (x1v1) (the two stability indices of x1v1 are drawn with red 
lines), which remains marginally stable, since we find that one of its stability indices 
stays on the $-2$ axis. 

We have taken several surfaces of section and we studied the structure of
the phase space at the neighborhood of the marginally stable p.o. x1v1. We 
first studied the phase space structure close to x1v1 for Ej=$-$4.4461, just
after the tangency of the b1 index with the b=$-2$ axis. We integrated 21
initial conditions at the neighborhood of x1v1 $(x_0,z_0,\dot{x_0},\dot{z_0})\approx$ 
$(1.298965,0.041883918,$\\ $-.0036186582,0.026639957)$, by 
perturbing the $x_0$ initial condition of x1v1  by  
$\Delta x_0 =  0.02,0.04\dots 0.42 $ successively. Fig.~\ref{proji2} gives the $(x,\dot x,z)$ 3D projections of the tori we found around x1v1. In Fig.~\ref{proji2}a we  
observe the first eight rotational tori $V_1, V_2, \dots V_8$  and in  
Fig. \ref{proji2}b the rotational tori $V_9 \dots V_{15}$. Finally 
in \ref{proji2}c we give the last six tori  
$V_{16} \dots V_{20}$ and $V_{21}$, of our sample. Each torus consists of 
$10^4$ consequents. In this dense coverage of the $\Delta x_0 =0.4$ space away from the $x_0$ initial condition of the x1v1 p.o., we encountered only rotational tori nested around x1v1, until we reached V21. There, perturbing $x_0$ by $\Delta x_0 =0.42$, the morphology of the torus changes. The V21 torus has a larger projection on the z-axis  and does not follow anymore the pattern of the nested tori.

We repeat the same procedure at a larger energy, Ej=$-$4.39. This time the $x$ initial  condition of x1v1 is perturbed successively by $\Delta x =  0.1,0.2,0.3$ and $0.4$. The result is given in Fig.~\ref{proji1}, having a point of view $(\theta, \phi) = (27^{o}, 316^{o})$. Four rotational  tori  surround the x1v1 p.o. and are indicated with arrows $(U_1, U_2, U_3$ and $U4)$. Each torus consists of $10^4$ consequents. The tori are distorted, with the last one having a morphology similar to V21 of the previous case.

\begin{figure}[h]
\begin{center}
\begin{tabular}{cc}
\resizebox{85mm}{!}{\includegraphics{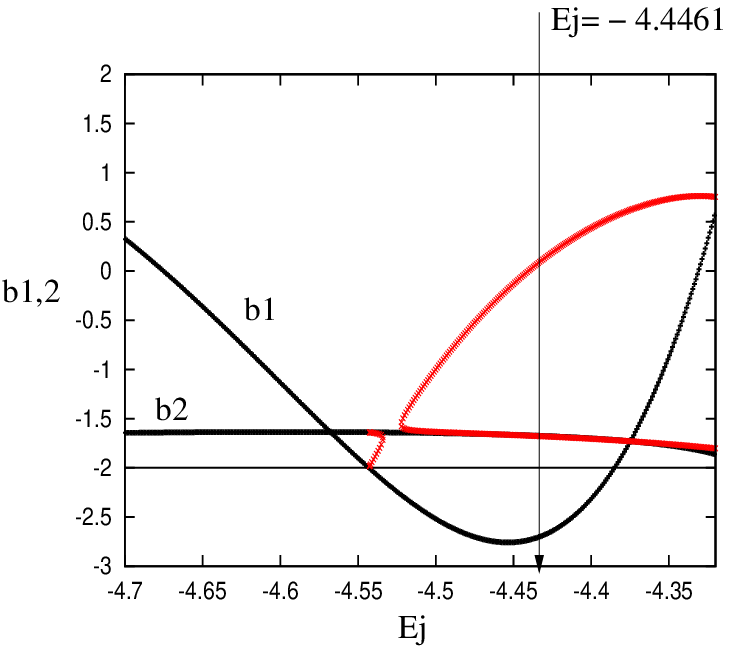}}\\
\end{tabular}
\caption{Stability Diagram  for $-4.7 < Ej < -4.3$, that shows the stability 
of the family x1 and its bifurcating family of p.o. x1v1.}
\label{stab2b}
\end{center}
\end{figure}

\begin{figure*}[t]
  \begin{center}
   \resizebox{120mm}{!}{\includegraphics{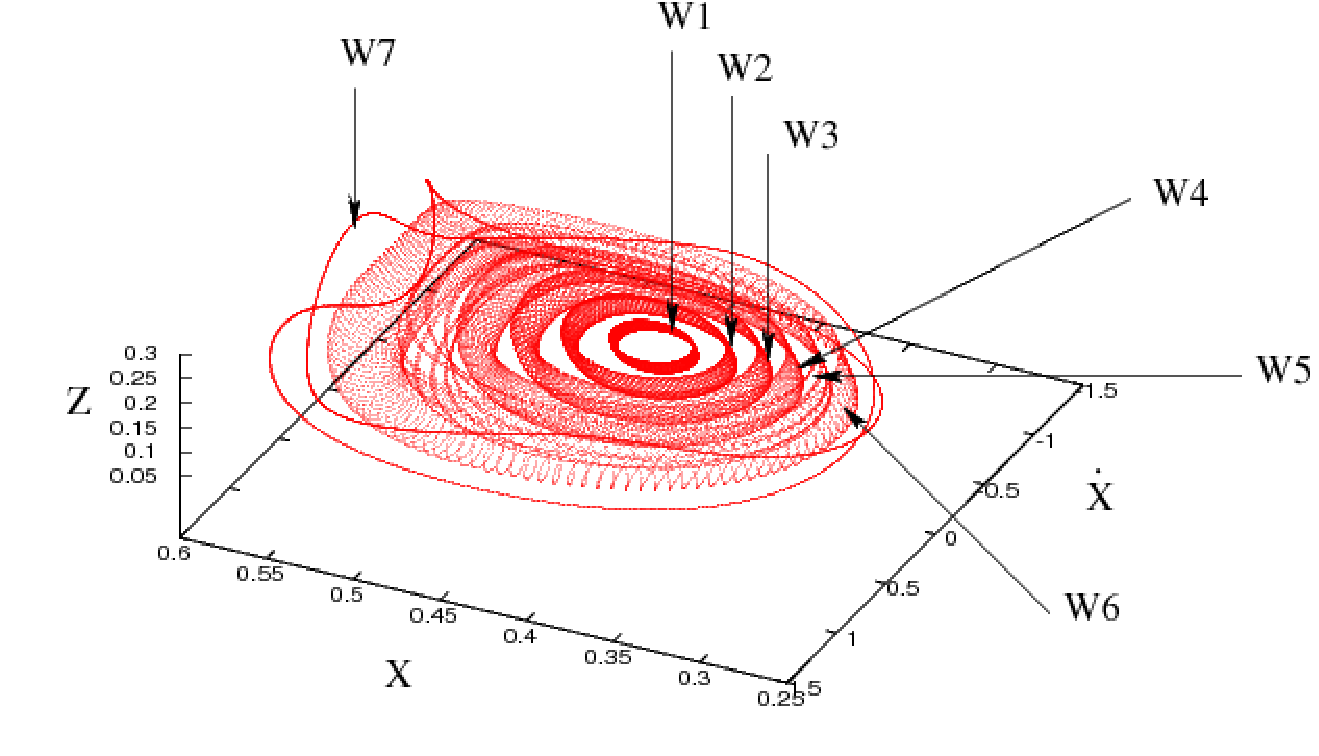}}\\
    \caption{The $(x,\dot x,z)$ 3D projection of the 
      4D surface of section for Ej=$-4.4461$ close to the stable p.o. x1v1. Our point of view
      is $(\theta, \phi) = (55^{o}, 194^{o})$.}
    \label{proji5}
  \end{center}
\end{figure*}

\subsection{The small perturbation case $q_a=1.01,q_b=1.0$}
In order to study qualitative differences from the axisymmetric case when a small perturbation is introduced in the system, we increase the $q_a$ parameter from $q_a=1$ to $q_a=1.01$.
Fig.~\ref{stab2b} gives for this system the evolution of the stability of the central family 
x1 and that of its bifurcations  for $-4.7 <$ Ej $< -4.3$. We observe that 
x1 (b1, b2 indices with black lines) is initially stable and at Ej=$-$4.54 
we have a transition from stability to simple instability. The family x1 becomes again stable at Ej=$-4.386$. At the transition from stability to simple instability x1v1 is bifurcated
(red lines show its stability indices ), as stable. At a larger energy (Ej=$-4.525$) we
have a $S\rightarrow \Delta$ transition to complex instability of the family x1v1. In this
case the stability indices become  complex numbers  and they do not appear in
the figure. There is a narrow complex unstable interval of x1v1 until Ej=$-4.525$, where x1v1 becomes again stable. In the present paper we study the phase space only in the neighborhood of stable p.o.

We choose a value of energy for which we have studied the orbital dynamics at the neighborhood of a stable x1v1 orbit at the axisymmetric case, Ej=$-$4.4461, and we perturb first he $x_0$ initial condition of x1v1 by $\Delta x =  0.02, 0.04$ \dots $0.14$ successively. 

In Fig.~\ref{proji5} we observe seven tori at the $(x,\dot x, z)$ projection, which we name $W_1, W_2,$\dots $W_7$,
surrounding the fixed point of the x1v1 p.o. at this $Ej$. Each torus 
consists of 10$^4$  consequents. The morphology of $W_1,W_2,W_3,W_4$ and $W_6$ indicates that they are 
rotational while the morphology of $W_5$ and $W_7$ is typical for tube tori. The tubes of $W_5$ have many self-intersections. However, the gaps between them do not fill even if we consider $10^6$ consequents. Thus, we conclude that $W_5$ is a tube torus. On the other hand, despite its wavy character, $W_7$ is a tube torus with only one self intersection in the 3D $(x,\dot x, z)$ projection. We note that tube tori appear in the 3D projections of the spaces of section as soon as a perturbation is introduced, even if it is a small one.

\section{Summary and Conclusions}
In this phenomenological paper we have studied in detail the structure of phase space in the neighborhood of stable periodic orbits in a 3D potential that represents a rotating, thick galactic disk. We have visualized the 4D spaces of section by means of the color and rotation method. This allowed us to clarify the properties of the invariant tori that we encounter in the vicinity of the stable p.o. We have also examined chaotic zones that we have found encaged between tori and the effect that these zones could have in the support of structures observed in thick galactic disks. Below we summarize our conclusions and compare our results with the results of previous studies.

\begin{enumerate}
\item \textit{On} the equatorial plane \textit{(z=0)}, the dynamics of the system is determined by the presence of the central family x1. Even in the intervals that x1 is characterized as simple unstable \textit{(U)}, we find in the $(x, \dot{x})$ surface of section invariant curves surrounding the periodic orbit. 
\item In general we find around stable periodic orbits invariant tori. In the
  \textit{3D projections} we encounter the invariant tori that have been found
  in the papers by Froeschl\'{e}  (1970, 1972), Martinet and Magnenat (1981), Contopoulos et al. (1982), Magnenat (1982), Patsis \& Zachilas (1994) and Vrahatis et al. (1997), named by the latter authors ``rotational tori''. However, we have found as well the other kind of invariant tori, named by Vrahatis et al. (1997) ``tube tori''. The fact that this second kind of tori appear frequently in the vicinity of stable periodic orbits in totally different physical systems as are Hamiltonian systems of galactic type (this paper) and 4D symplectic maps associated with the problem of beam stability in circular particle accelerators (Vrahatis et al. 1997), indicates that they represent a generic behavior in 3D orbits.
\item The method of color and rotation was particularly enlightening in understanding the properties of the two kinds of tori. 
\begin{itemize}
 \item As regards the \textit{rotational tori} in all cases we studied in our potential the continuation of color, that guarantees the smooth variation of the 4th coordinate, changes side along certain lines (Figs.~\ref{rot1}, \ref{rot2}). This shows that \textit{in the 3D projections} we have intersections in the rotational tori, which, in all cases we encountered in this study, appear four times on the tori.
\item As regards the \textit{tube tori}, their characteristic feature, the self-intersections of their tubes, was a projection effect in all cases, since at the intersections meet always different colors. We remark that we found always an odd number of self-intersections in the 3D projections of the tube tori. \textit{In the 4D space} we did not find any self-intersections of the tube tori.
\end{itemize}
We found that the two kinds of tori cannot be distinguished neither from their rotation numbers, nor from the morphological features of the orbits in the configuration space.
\item As we depart from the initial conditions of the stable p.o., the 3D projections of the tori appear distorted. Beyond a certain torus we found sticky chaotic orbits that for a long time remain on toroidal surfaces on which the consequents have a smooth color variation and finally diffuse in phase space, where we observe mixing of colors.
\item We found chaotic orbits, which can be encaged by tori in the 4D space for times longer than a Hubble time, but not forever, since there are no topological barriers. We presented an example, where such orbits support a peanut morphology, similar but thicker than the morphology supported by the corresponding quasi-periodic orbits trapped around the stable x1v1.
\item Tube-tori have been found only in the 3D projections of orbits in perturbed systems. In the axisymmetric case we encountered only rotational tori.
\end{enumerate}

\vspace{2cm}
\textit{Acknowledgments} We acknowledge many stimulating discussions and very useful comments from Prof. G.~Contopoulos. We also acknowledge fruitful discussions with Prof. A. Pinotsis, Prof. M. Vrahatis and Dr. Ch. Skokos. This research has been partly supported by the Research Committee of the Academy of Athens under the project 200/739.\\

\section{References} 
 Abraham R. and  Marsden J.E. [1978] \textit{Foundations of Mechanics}, 
 Benjamin-Cummings Publ. Co., Reading, Massachusetts.\\
  Arnold V.I. and  Givental A.B. [2000] ``Symplectic Geometry''  In: 
  \textit{Dynamical Systems IV}, ed  by  V.I. Arnold, S.P. Novikov, 
  Springer-Verlag, New York Berlin Heidelberg, pp 1--138.\\
  Arnold V.I. [1963] ``Proof of a theorem of A.N.~Kolmogorov on the invariance 
  of quasi-periodic motions under small perturbations of the Hamiltonian'' 
  \textit{Russ. Math. Surveys} \textbf{18}, 9-36.\\ 
  Binney J. and  Tremaine S. [2008] \textit{Galactic  Dynamics}, 
  Princeton Univ. Press, Princeton.\\
  Broucke R. [1969] ``Periodic Orbits in the Elliptic  Restricted  Three-Body 
  Problem'' \textit{NASA Tech. Rep.} 32-1360, 1-125.\\
  Contopoulos G. [2002] \textit{Order and Chaos in Dynamical Astronomy},
  Springer-Verlag, New York Berlin Heidelberg.\\
  Contopoulos G. and Papayannopoulos Th. [1980] ``Orbits in weak and strong 
  bars'' \textit{Astron. Astrophys.} \textbf{92}, 33-46.\\
  Contopoulos G. and Barbanis B. [1985] ``Resonant systems with three degrees 
  of freedom''  \textit{Astron. Astrophys.} \textbf{153}, 44-54.\\
  Contopoulos G. and Harsoula M. [2008] ``Stickiness in Chaos''
  \textit{Int. J. Bif. Chaos} \textbf{18}, 2929-2949.\\
  Contopoulos G. and Patsis P.A. [2006] ``Outer dynamics and escapes in barred
  galaxies'' \textit{Mon. Not. R. Astr. Soc.} \textbf{369}, 1039-1054.\\
  Contopoulos G. and Magnenat P. [1985] ``Simple three-dimensional  periodic 
  orbits in a galactic-type potential'' \textit{Celest. Mech.} \textbf{37}, 
  387-414. \\
  Contopoulos G., Magnenat P. and Martinet L. [1982] ``Invariant surfaces and 
  orbital behavior in dynamical systems of 3 degrees of freedom II'' 
  \textit{Physica D} \textbf{6}, 123-136. \\
  Debattista V., Mayer L., Carollo C.M., Moore B., Wadsley J. and Quinn T. 
  [2006] ``The secular evolution of disk structural parameters'',
  \textit{Astrophys. J}  \textbf{645}, 209-227. \\
  Englmaier P. and Gerhard O.E. [1999]  ``Gas dynamics and large-scale 
  morphology of the Milky Way  galaxy'' \textit{Mon. Not. R. Astr. Soc.} 
  \textbf{304}, 512-534.\\
  Froeschl\'{e} C. [1970] ``Numerical study of dynamical systems with three 
  degrees of freedom'' \textit{Astron. Astrophys.} \textbf{4}, 115-128.\\
  Froeschl\'{e} C. [1972] `` Numerical study of a four-dimensional mapping'' 
  \textit{Astron. Astrophys.} \textbf{16}, 172-189.\\
  Hadjidemetriou J. [1975] ``The stability of periodic orbits in the 
  three-body problem'' \textit{Celest. Mech.}  \textbf{ 12}, 255-276.\\
  Kolmogorov A.N. [1954] ``On  the conservation of 
    conditionally  periodic motions  under small perturbations of the 
    Hamiltonian'' \textit{Dokl. Akad. Nauk USSR} \textbf{98}, 527-530.\\
  Kuksin S. and P$\ddot o$schel J. [1994] ``On the Inclusion of Analytic
  Symplectic Maps in Analytic Hamiltonian Flows and its applications'' In: 
  \textit{Seminars on Dynamical Systems}, ed by  S. Kuksin, V. Lazutkin, 
  J. P$\ddot o$schel, Birkh$\ddot a$user: Basel.\\
  Lichtenberg A.J. and Lieberman M.A. [1992]  \textit{Regular and 
    Chaotic Dynamics}, Springer-Verlag, Berlin Heidelberg New York.\\
  Magnenat P. [1982] ``Numerical study of periodic orbit properties in a 
  dynamical system  with three  degrees of freedom''  
  \textit{Celest. Mech.} \textbf{28}, 319-343.\\   
  Martinet L. and Magnenat P. [1981] `` Invariant surfaces and orbital 
  behavior in  dynamical systems with 3 degrees of freedom.'' \textit{Astron. 
    Astrophys.} \textbf{96}, 68-77.\\
  Miyamoto M. and  Nagai  R. [1975] ``Three-dimensional models for the 
  distribution of mass in galaxies'' \textit{Publ. Astron. Soc. Japan} 
  \textbf{27}, 533-543.\\ 
  Moser J. [1962] ``On invariant curves of an area preserving mappings
  of an annulus''\textit{ Nachr. Akad. Wiss. G$\ddot o$tt, II Math.- Phys Kl.}
  1-20.\\
  Patsis P.A. and Grosb{\o}l P. [1996] ``Thick spirals: dynamics and orbital 
  behavior'' \textit{Astron. Astrophys.} \textbf{315}, 371-383.\\
  Patsis P.A. and Zachilas L. [1994] `` Using Color and rotation for 
  visualizing  four-dimensional Poincar\'{e}  cross-sections: with 
  applications to the  orbital  behavior of a three-dimensional 
  Hamiltonian system'' \textit{Int. J. Bif. Chaos} \textbf{ 4}, 1399-1424.\\
  Patsis P.A., Skokos Ch. and Athanassoula E. [2002] ``Orbital dynamics of
  three-dimensional bars-III. Boxy/peanut edge-on profiles'' 
  \textit{Mon Not. R. Astr. Soc.} \textbf{337}, 578-596.\\
  Pfenniger D. [1985a] `` Numerical study of complex instability:I Mappings''
  \textit{Astron. Astrophys} \textbf{150}, 97-111.\\
  Pfenniger D. [1985b], ``Numerical study of complex instability:II Barred 
  galaxy bulges'' \textit{Astron. Astrophys.} \textbf{150}, 112-128.\\
  Poincar\'{e} H. [1892] ``Les M\'{e}thodes  Nouvelles  de la  
  M\'{e}canique  C\'{e}leste'' Gauthier Villars, Paris I (1892), II (1893), 
  III (1899); Dover (1957).\\
  Skokos Ch., Contopoulos G. and Polymilis C. [1997] ``Structures in the phase
  space of a four dimensional symplectic map'' \textit{Celest. Mech. 
    Dyn. Astron.} \textbf{65}, 223-251.\\
  Skokos Ch., Contopoulos G. and Polymilis C. [1999] ``Numerical study of the
  phase space of a four dimensional symplectic map'' In \textit{Hamiltonian
    Systems with three or more  degrees of freedom}, ed. by Sim\'{o} C., 
  Plenum Press, p. 583-587. \\
  Skokos Ch. [2010] ``The Lyapunov characteristic exponents and their 
  computation''  \textit{Lect. Notes Phys.} \textbf{790}, 63-135.\\
  Skokos Ch. [2001] ``On the stability of periodic orbits of high dimensional 
  autonomous Hamiltonian systems'' \textit{Physica D} \textbf{159}, 155-179.\\
  Skokos Ch., Patsis P.A. and  Athanassoula E. [2002a] `` Orbital dynamics 
  of three-dimensional bars-I. The backbone of three-dimensional bars. 
  A fiducial case'' \textit{Mon. Not. R. Astr. Soc.} \textbf{333}, 847-860.\\
  Skokos  Ch., Patsis P.A. and  Athanassoula E. [2002b] ``Orbital dynamics of 
  three-dimensional bars-II. Investigation of the parameter space''  
  \textit{Mon. Not. R. Astr. Soc.} \textbf {333}, 861-870.\\
  Vrahatis M.N., Bountis T.C. and Kollmann M. [1996] ``Periodic orbits and
  invariant surfaces of 4D nonlinear mappings'' \textit{Int. J. Bif. Chaos} 
  \textbf{6}, 1425-1437.\\ 
  Vrahatis M.N., Isliker H. and Bountis T.C. [1997] ``Structure and breakdown
  of invariant tori in a 4-D mapping model of accelerator dynamics''
  \textit{Int. J. Bif. Chaos} \textbf{7}, 2707-2722.\\
  Wiggins S. [2003] \textit{Introduction to Applied  Nonlinear Dynamical 
    Systems  and Chaos}, Springer-Verlag, Berlin Heidelberg New York.\\
  Wolfram S. [1999] \textit{The Mathematica book}, Wolfram media \&
  Cambridge  Univ. Press.

\end{document}